%% file: IIB-AdS3-S-matrix-arXiv-v2.tex
\documentclass[a4paper,11pt]{article}
\usepackage{a4wide,amsmath,amssymb,bbm}
\usepackage[english]{babel}

\allowdisplaybreaks[2]

\pdfoutput=1

\makeatletter

\@addtoreset{equation}{section} \makeatother

\usepackage{titletoc}
\dottedcontents{section}[6mm]{\smallskip }{6mm}{0pc}
\dottedcontents{subsection}[8mm]{\itshape\small }{6mm}{0pc}
\dottedcontents{subsubsection}[16mm]{\itshape\small }{10mm}{0pc}
\usepackage[nosort]{cite}

\usepackage[parsep]{collref}

\usepackage{graphicx}
\usepackage{feynmp}
\DeclareGraphicsExtensions{.pdf}  % include pdf files by default

\def \bea  {\begin{eqnarray}}
\def \eea  {\end{eqnarray}}

\newcommand{\ket}[1]{|{#1}\rangle}
\newcommand{\nn}{\nonumber}

\usepackage{parskip}

\def\unit{\protect{{1 \kern-.28em {\rm l}}}}

\begin{document}
%% Feynman diagrams
\input{dgms.tex}

\begin{flushright}
Imperial-TP-LW-2016-01\\
ZMP-HH/16-9.
\end{flushright}

\vspace{50pt}

\begin{center}
{\LARGE{\bf The complete one-loop BMN S-matrix in $AdS_3\times S^3\times T^4$}}

\vspace{50pt}

{\bf \large  Per Sundin$^a$} and {\bf\large Linus Wulff$\, ^b$}

\vspace{15pt}

{$^a$ \it\small II. Institut f\"ur Theoretische Physik, Universit\"at Hamburg, \\
Luruper Chaussee 149, 22761 Hamburg, Germany}\\
\vspace{5pt}
{$^b$ \it\small Blackett Laboratory, Imperial College, London SW7 2AZ, U.K.}\\

\vspace{90pt}

{\bf Abstract}

\end{center}
\noindent
We compute the full one-loop 2-particle S-matrix for excitations of the type IIB $AdS_3\times S^3\times T^4$ BMN string. The S-matrix is found to respect the expected symmetries and the phases are consistent with the crossing equations. By analyzing how the relevant integrals scale with the IR regulator we show that scattering of massless bosons is trivial at two loops. Based on our results we argue that the additional $\mathfrak{su}(2)$ S-matrix appearing in the massless sector in the exact solution should trivialize.

\pagebreak 
\tableofcontents
\setcounter{page}{1}

%%%%%%%%%%%%%%%%%%%%%%%%%%%%%%%%%%%%%%%%%%%%%%%%%%%%%%%%%%%%%%%%%%%%%%%%

\section{Introduction}
String theory in $AdS_3\times S^3\times T^4$ preserving 16 supersymmetries is classically integrable\footnote{This is a general property of superstrings in symmetric space backgrounds that preserve some supersymmetry \cite{Wulff:2015mwa}.} \cite{Babichenko:2009dk,Sundin:2012gc,Cagnazzo:2012se,Sundin:2013uca,Wulff:2014kja} and is expected to be integrable also at the quantum level. If this is indeed the case it should be possible to determine the spectrum along similar lines as for the string in $AdS_5\times S^5$ \cite{Beisert:2010jr}. Here we will consider the type IIB $AdS_3\times S^3\times T^4$ solution supported by Ramond-Ramond three-form flux. An important first step in solving the model is to find the exact S-matrix. Up to phase factors the S-matrix can be completely determined by the symmetries \cite{Borsato:2014hja}. Since the BMN string has both massive and massless excitations the (2-particle) S-matrix splits into three sectors: massive-massive scattering, massive-massless scattering and massless-massless scattering. Each sector comes with phase factors which should be determined by crossing symmetry together with some input from perturbation theory. In the massive sector there are two phases whose all-loop crossing symmetric solution was conjectured in \cite{Borsato:2013hoa} and agrees with perturbative worldsheet calculations at one and two loops \cite{Sundin:2013ypa,Engelund:2013fja,Abbott:2013mpa,Bianchi:2014rfa,Roiban:2014cia}.

Here we complete the one-loop analysis of the S-matrix by computing the full one-loop 2-particle S-matrix including mixed and massless sectors. The structure of the S-matrix is found to agree with that of \cite{Borsato:2014hja}\footnote{A change in the normalization of the mixed sector S-matrix of that paper is needed for agreement.} and the one-loop phases we find in the mixed and massless sectors are shown to respect crossing symmetry. In the mixed sector we show that one-loop scattering is consistent with integrability by verifying that the amplitudes for scattering processes where the particles change their momenta vanish. We are further able to show that the scattering of massless bosons is trivial at two loops, as was argued for a particular process in the type IIA setting in \cite{Sundin:2015uva}. Our perturbative calculations suggest that the extra $\mathfrak{su}(2)$ S-matrix appearing in the massless sector in \cite{Borsato:2014hja} trivializes. The perturbative findings we present in this paper were recently used in \cite{Borsato:2016kbm} to conjecture all-loop expressions for the mixed and massless sector phases together with the corresponding Bethe equations.\footnote{For some reservations about the usefulness of said equations see \cite{Abbott:2015pps}.} 

In addition to the S-matrix the exact dispersion relation for the massive and massless modes is also determined by the symmetry analysis of \cite{Borsato:2014hja}. For the massive bosons the dispersion relation agrees with perturbation theory up to two loops, as earlier shown in the type IIA setting in \cite{Sundin:2014ema,Sundin:2015uva}. However, there a discrepancy in the two-loop correction for the massless bosons was found. Here we repeat these calculations for the type IIB string and also include the fermionic modes finding the same result. While our findings are consistent with worldsheet supersymmetry to the two loop order an explanation for the mismatch is still lacking. 

The outline of the paper is as follows. In section \ref{sec:GS} we describe the Green-Schwarz string action in $AdS_3\times S^3\times T^4$ to quartic order in fermions, its light-cone gauge-fixing and BMN expansion. In section \ref{sec:regularization} we describe our regularization scheme for the relevant one-loop integrals. Section \ref{sec:dispersion} gives the results of the two-loop two-point function calculations and correction to the dispersion relations. In section \ref{sec:S-matrix} we give the form of the S-matrix and section \ref{sec:phases} contains a discussion of the mixed and massless sector phases and crossing symmetry. We end with some conclusions and an appendix containing some details of the basic one-loop bubble integrals.

\section{Green-Schwarz string in $AdS_3\times S^3\times T^4$ with RR flux}
\label{sec:GS}
The Green-Schwarz superstring action can be expanded order by order in fermions as
\begin{equation}
S=g\int d^2\xi\,(\mathcal L^{(0)}+\mathcal L^{(2)}+\ldots)\,,
\end{equation}
where we let $g$ denote the string tension. In a general type II supergravity background the form of this expansion is known explicitly up to quartic order in fermions \cite{Wulff:2013kga}. This is the action we will use for the string in type IIB $AdS_3\times S^3\times T^4$. The purely bosonic terms in the Lagrangian are given by
\begin{equation}
\mathcal L^{(0)}=\frac12\gamma^{ij}e_i{}^ae_j{}^b\eta_{ab}\,,\qquad\gamma^{ij}=\sqrt{-h}h^{ij}\,,
\label{eq:L0}
\end{equation}
where $e^a$ $(a=0,\ldots,9)$ are the (bosonic) vielbeins and we used the fact that the bosonic B-field vanishes in our case. The vielbeins can be read off from the metric which we take to be
\begin{align}
ds^2_{AdS_3}=-\left(\frac{1+\frac12|z|^2}{1-\frac12|z|^2}\right)^2dt^2+\frac{2|dz|^2}{(1-\frac12|z|^2)^2}\,,
\qquad
ds^2_{S^3}=\left(\frac{1-\frac12|y|^2}{1+\frac12|y|^2}\right)^2d\varphi^2+\frac{2|dy|^2}{(1+\frac12|y|^2)^2}\,,
\label{eq:metric}
\end{align}
for complex transverse coordinates $z,y$ while the metric of $T^4$ is $2|du_1|^2+2|du_2|^2$.

The terms quadratic in fermions take the form
\begin{equation}
\mathcal L^{(2)}=ie_i{}^a\,\theta^I\gamma_aK_{IJ}^{ij}\mathcal D^{JK}_j\theta^K\,,\qquad K^{ij}_{IJ}=\delta_{IJ}\gamma^{ij}-\varepsilon^{ij}\sigma^3_{IJ}\,,
\end{equation}
where $\theta^I$ ($I=1,2$) are the two 16-component Majorana-Weyl spinors of type IIB and the Killing spinor derivative operator is given by
\begin{equation}
\label{eq:DbA}
\mathcal D_i\theta^I=
\big(\delta^{IJ}\partial_i-\frac{1}{4}\delta^{IJ}\omega_i{}^{ab}\gamma_{ab}-\frac{1}{2}e_i{}^a\sigma^1_{IJ}\mathcal{P}_8\gamma^{012}\gamma_a\big)\theta^J\,,
\qquad\mathcal{P}_8=\frac{1}{2}\big(1+\gamma^{012345}\big)\,,
\end{equation}
where $\omega^{ab}$ is the spin connection and $\mathcal P_8$ projects on the $2\times8$ supersymmetric fermionic directions. The last term comes from the coupling to the RR three-form flux $F^{(3)}\simeq\Omega_{AdS_3}+\Omega_{S^3}$.

Finally the quartic terms in the Lagrangian take the form (suppressing the $SO(2)$-indices)
\begin{align}
\mathcal L^{(4)}=&
-\frac{1}{2}\theta\gamma^a\mathcal D_i\theta\,\theta\gamma_aK^{ij}\mathcal D_j\theta
+\frac{i}{6}e_i{}^a\,\theta\gamma_aK^{ij}\mathcal M\mathcal D_j\theta
-\frac{i}{12}e_i{}^ae_j{}^b\,\theta\gamma_aK^{ij}(M+\sigma^3M\sigma^3)\sigma^1\mathcal{P}_8\gamma^{012}\gamma_b\theta
\nonumber\\
&{}
+\frac{1}{16}e_i{}^ce_j{}^d\,\theta\gamma_c{}^{ab}K^{ij}\theta\,\theta\Gamma_dU_{ab}\upsilon
-\frac{1}{16}e_i{}^ce_j{}^d\,\theta\gamma_c{}^{ab}\sigma^3K^{ij}\theta\,\theta\gamma_d\sigma^3U_{ab}\upsilon
\nonumber\\
&{}
-\frac{1}{24}e_i{}^ce_j{}^d\,\theta\gamma_c{}^{ab}K^{ij}\theta\,\theta\gamma_aU_{bd}\upsilon
-\frac{1}{24}e_i{}^ce_j{}^d\,\theta\gamma_c{}^{ab}\sigma^3K^{ij}\theta\,\theta\gamma_a\sigma^3U_{bd}\upsilon\,,
\label{eq:L4}
\end{align}
where 
\begin{equation}
\upsilon=(1-\mathcal P_8)\theta\,,\qquad U_{ab}=\frac{1}{2}\mathcal{P}_8\gamma^{012}\gamma_{[a}\mathcal{P}_8\gamma^{012}\gamma_{b]}-\frac{1}{4}R_{ab}{}^{cd}\,\gamma_{cd}\,,
\end{equation}
and
\begin{align}
\mathcal M^{\alpha I}{}_{\beta J}=&{}
M^{\alpha I}{}_{\beta J}
+(\sigma^3M\sigma^3)^{\alpha I}{}_{\beta J}
-\frac{i}{2}(\sigma^1\mathcal{P}_8\gamma^{012}\gamma_a\theta)^{\alpha I}\,(\theta\gamma^a)_{\beta J}
+\frac{i}{4}(\gamma^{ab}\theta)^{\alpha I}\,(\theta\gamma_a\mathcal{P}_8\gamma^{012}\gamma_b\sigma^1)_{\beta J}\,,
\nonumber\\
M^{\alpha I}{}_{\beta J}=&{}
\frac{i}{2}\upsilon\sigma^1\gamma^{012}\upsilon\,\delta^\alpha_\beta\delta_{IJ}
-i\theta^{\alpha I}\, (\upsilon\gamma^{012}\sigma^1)_{\beta J}
+i(\sigma^1\gamma^a\gamma^{012}\upsilon)^{\alpha I}\,(\theta\gamma_a)_{\beta J}\,.
\label{eq:M}
%
%S=-4\sigma^1_{IJ}\mathcal{P}_8\gamma^{012}
\end{align}

Equipped with this string action the next step is to expand around the point-like BMN string solution given by  \cite{Berenstein:2002jq}
\begin{equation}
x^+=\frac12(t+\varphi)=\tau\,.
\end{equation}
At the same time we fix the light-cone gauge and corresponding kappa symmetry gauge
\begin{equation}
\label{eq:gf}
x^+=\tau\,,\qquad\gamma^+\theta^I=0\,.
\end{equation}
The Virasoro constraints can then be used to solve for $x^-$ in terms of the other fields. In this gauge the worldsheet metric defined in (\ref{eq:L0}) takes the form $\gamma=\eta+\hat\gamma$, where $\hat\gamma$ denotes higher order corrections determined by the following conditions on the momentum conjugate to $x^-$
\begin{equation}
\label{eq:gf1}
p^+=-\frac12\frac{\partial\mathcal L}{\partial \dot x^-}=1\,,\qquad\frac{\partial\mathcal L}{\partial x^-{}'}=0\,.
\end{equation}
This is often referred to as uniform light-cone gauge.

Rescaling the transverse fields by a factor $g^{-1/2}$ yields a perturbative expansion in the string tension\footnote{The fact that only even orders appear in this expansion for $AdS\times S\times T$ backgrounds greatly simplifies a perturbative treatment.}
\bea \nn
\mathcal{L}= \mathcal{L}_2+\frac{1}{g}\mathcal{L}_4+\frac{1}{g^2}\mathcal{L}_6+\dots
\eea
where the subscript denotes the number of transverse coordinates. The quadratic Lagrangian takes the form ($\partial_\pm=\partial_0\pm\partial_1$)
\begin{align}
\mathcal{L}_2=&
|\partial z|^2
-|z|^2
+|\partial y|^2
-|y|^2
+|\partial u_1|^2
+|\partial u_2|^2
+i\bar\chi_L^1\partial_-\chi_L^1
+i\bar\chi_R^1\partial_+\chi_R^1
-\bar\chi_L^1\chi_R^1
-\bar\chi_R^1\chi_L^1
\nonumber\\
&{}
+i\bar\chi_L^2\partial_-\chi_L^2
+i\bar\chi_R^2\partial_+\chi_R^2
-\bar\chi_L^2\chi_R^2
-\bar\chi_R^2\chi_L^2
+i\bar\chi_L^3\partial_-\chi_L^3
+i\bar\chi_R^3\partial_+\chi_R^3
+i\bar\chi_L^4\partial_-\chi_L^4
+i\bar\chi_R^4\partial_+\chi_R^4\,.
\label{eq:quadratic-L}
\end{align}
The spectrum consists of 4+4 massive and 4+4 massless modes. They are charged under four $U(1)$'s and the charges are given in table \ref{tab:charges-ads3}. The interaction terms in the Lagrangian are quite complicated and we will not give them here since they are straight-forwardly found by expanding the original Lagrangian.

\begin{table}[ht]
\begin{center}
\begin{tabular}{ccccccccc}
& $z$ & $y$ & $u_1$ & $u_2$ & $\chi^1$ & $\chi^2$ & $\chi^3$ & $\chi^4$ \\
\hline
\vphantom{${}^{A^{A^a}}$}
$U(1)_1$ & $-1$ & $0$ & $0$ & $0$ & $-1/2$ & $1/2$ & $1/2$ & $1/2$  \\
$U(1)_2$ & $0$ & $-1$ & $0$ & $0$ & $1/2$ & $-1/2$ & $1/2$ & $1/2$ \\
$U(1)_3$ & $0$ & $0$ & $-1$ & $0$ & $1/2$ & $1/2$ & $-1/2$ & $1/2$ \\
$U(1)_4$ & $0$ & $0$ & $0$ & $-1$ & $1/2$ & $1/2$ & $1/2$ & $-1/2$
\end{tabular}
\end{center}
\caption{Summary of $U(1)$ charges for $AdS_3\times S^3\times T^4$. }
\label{tab:charges-ads3}
\end{table}

\section{One-loop regularization}\label{sec:regularization}
At one loop one encounters two types of integrals: Bubble integrals
\begin{equation}
B^{rs}_{m_1m_2}(P)=\int\frac{d^2k}{(2\pi)^2}\frac{k_+^r k_-^s}{(k^2-m_1^2)((k-P)^2-m_2^2)}\,,
\label{eq:bubble-int}
\end{equation}
coming from the diagrams in (\ref{diagram:stu}), and Tadpole integrals
\bea
T^{rs}(P) =\int\frac{d^2k}{(2\pi)^2}\frac{k_+^r k_-^s}{(k-P)^2-m^2}\,,
\eea 
coming from the diagrams in (\ref{diagram:t6}). Here $P$ is a combination of external momenta. Many of these are power-counting UV divergent,\footnote{Due to the presence of massless modes there are also IR divergences. These are dealt with by introducing a small regulator mass $\mu$ for the massless modes which is taken to zero at the end.} eg. $B^{rs}$ for $r+s>1$ and need to be regularized. In \cite{Roiban:2014cia} it was shown that standard dimensional regularization gives the wrong answer. It leads to an S-matrix not compatible with the symmetries due to the introduction of extra rational terms. In the same paper a different regularization scheme, essentially using tensor reduction in strictly two dimensions instead, was introduced which was shown to give a result consistent with the symmetries of the BMN vacuum. The same procedure has been shown to give the right answer also at two loops \cite{Sundin:2014ema} as well as for the spinning GKP string \cite{Bianchi:2015vgw}.

The idea is to use the algebraic identity
\begin{equation}
\frac{1}{(k-P)^2-m_2^2}-\frac{1}{k^2-m_1^2} = \frac{k_+ P_- + k_- P_+-P^2+m_2^2-m_1^2}{(k^2-m_1^2)((k-P)^2-m_2^2)}
\end{equation}
to reduce all bubble integrals to the (UV) finite integral $B^{00}$ plus tadpole integrals. Once all divergences are isolated in tadpole integrals these can be evaluated in dimensional regularization without introducing unwanted rational terms. Multiplying the above identity by $k_+^rk_-^s$ and integrating we find\footnote{We will suppress the $m_1m_2$ subscript on the bubble integrals when there is no risk of confusion.}
\begin{equation}
\label{eq:bubble-id1}
P_-B^{r+1,s}(P)+P_+B^{r,s+1}(P)+(m_2^2-m_1^2-P^2)B^{rs}(P)= T^{rs}_{[m_2]}(P)-T^{rs}_{[m_1]}(0)\,,
\end{equation}
where we have indicated explicitly the mass in the tadpole integrals. We also use the very simple fact that for $r,s\geq 1$
\begin{equation}
B^{rs}(P)= m_1^2 B^{r-1,s-1}(P)+T^{r-1,s-1}_{[m_2]}(P)\,,
\label{eq:bubble-id2}
\end{equation}
which follows by writing $k_+k_-=k^2=k^2-m^2+m^2$ and simplifying the integrand. Finally $B^{10}$ and $B^{01}$, which are UV finite, can easily be computed by the following trick. First we note that $B^{10}(P)$ and $B^{01}(P)$ are odd functions of $P$ as follows by shifting the integration variable. But since $P_-B^{10}(P)+P_+B^{01}(P)$ is Lorentz invariant they must in fact be of the form 
\begin{equation}
B^{10}(P)=P_+f(P^2,m_1,m_2)\,,\qquad B^{01}(P)=P_-f(P^2,m_1,m_2)\,,
\end{equation}
where $f$ is some function that must be the same for the two integrals by symmetry. Using this in (\ref{eq:bubble-id1}) with $r=s=0$ we can determine the function $f$ and finally we find\footnote{We have shifted the integration variable in the first tadpole integral.}
\bea
&&B^{10}(P)=\frac{P_+}{2P^2}[(P^2+m_1^2-m_2^2)B^{00}(P)+T^{00}_{[m_2]}(0)-T^{00}_{[m_1]}(0)]\,,\\
&&B^{01}(P)=\frac{P_-}{2P^2}[(P^2+m_1^2-m_2^2)B^{00}(P)+T^{00}_{[m_2]}(0)-T^{00}_{[m_1]}(0)]\,.
\eea
Using (\ref{eq:bubble-id1}) and (\ref{eq:bubble-id2}) we also find (again we have used shifts of the integration variable and Lorentz invariance to simplify the tadpole contributions)
\bea 
&& B^{30}(P)=-\frac{1}{P_-}[(m_2^2-m_1^2-P^2)B^{20}(P)+P_+m_1^2 B^{10}(P)]\,,\\
&& B^{03}(P)=-\frac{1}{P_+}[(m_2^2-m_1^2-P^2)B^{02}(P)+P_-m_1^2 B^{01}(P)]\,,\\
&& B^{20}(P)=-\frac{1}{P_-}[(m_2^2-m_1^2-P^2)B^{10}(P)+P_+m_1^2 B^{00}(P)]\,,\\
&& B^{02}(P)=-\frac{1}{P_+}[(m_2^2-m_1^2-P^2)B^{01}(P)+P_-m_1^2 B^{00}(P)]\,.
\eea
These are all the relations we will need for our one loop calculations.\footnote{
These integral identities can also be used to compute the one loop correction to the two-point function for the BMN modes in $AdS_3 \times S^3 \times S^3 \times S^1$ originally done in \cite{Sundin:2012gc}. In contrast to the $T^4$ case the excitations now have four different masses $(1,\alpha,1-\alpha,0)$ where $\alpha$ controls the relative size of the two $S^3$-factors and lies in the interval $0 \leq \alpha \leq 1$. For the lighter excitations with mass $\alpha$ and $1-\alpha$ one finds,
\bea \nn
\langle \bar y_i y_i \rangle^{(1)} = \langle \chi_i \chi_i \rangle^{(1)} = -2 p_1^2 \big(T_1^{00}(0)-\alpha T^{00}_\alpha(0)- (1-\alpha) T^{00}_{1-\alpha}(0)\big)
=\frac{i}{\pi}\big(\alpha\log\alpha+(1-\alpha)\log(1-\alpha)\big)p_1^2
\,,
\eea
where we used dimensional regularization in the last step. If desired one can choose a different regulator for tadpole integrals with different masses in the loop. While not very natural from the point of view of the sigma model, it is possible to make the entire one-loop contribution vanish in this way. 

The exact solution indicates that the heavy mode from $AdS_3$ of mass 1 should be a composite state made out of two lighter ones. The way this can happen in the one-loop worldsheet theory is if the bubble contribution to the propagator replaces the pole in the propagator with a branch cut \cite{Zarembo:2009au}. However, in the regularization outlined above we find
\bea \nn
\langle \bar y_1 y_1 \rangle^{(1)} = -\frac{1}{2}(1-\alpha)\alpha (p_+-p_-)^2\big(p^2-1\big)B_{\alpha,1-\alpha}^{00}+\dots
\eea
where the dots denote tadpole contributions. Once we go on-shell, $p^2=1$, the entire bubble contribution vanishes so we are lead to the conclusion that in order to see any indication of a composite state we need to go to higher loop order.
} The integrated result for the basic bubble integral $B^{00}$ is presented in appendix \ref{sec:appendix-integrals}.

\section{Two-point functions and two-loop correction to dispersion relations}\label{sec:dispersion}
To be able to calculate the 2-particle S-matrix we must first compute the off-shell one-loop two-point functions. One finds that only the massive bosons receive a correction which takes the form of wave function renormalization \cite{Roiban:2014cia} 
\bea
\langle \bar z z\rangle = \frac{i Z_z}{p^2-1}+\mathcal{O}(g^{-2})\,, \qquad
\langle \bar y y\rangle = \frac{i Z_y}{p^2-1}+\mathcal{O}(g^{-2})\,,
\eea
with the renormalization factors given by
\bea
Z_z = 1 +\frac{1}{4 \pi}\frac{1}{g} \big( -\frac{2}{\epsilon} + \log \pi + \gamma_E\big)\,, \qquad
Z_y = 1 -\frac{1}{4 \pi}\frac{1}{g} \big( -\frac{2}{\epsilon} + \log \pi + \gamma_E\big)\,,
\label{eq:Zs}
\eea
where the contributing tadpole integral has been evaluated in dimensional regularization in $d=2-\epsilon$ dimensions. This wave function renormalization must be taken into account to get a finite S-matrix.

Before we move on to the S-matrix however we will compute the two-loop correction to the on-shell two-point function, i.e. to the dispersion relation. This was done for bosonic modes in \cite{Sundin:2014ema, Sundin:2015uva} (in the type IIA context) and here we will extend that analysis to include also the fermionic modes. 

The perturbative result should be compared with the exact dispersion relation fixed by the underlying symmetries. For an excitation of mass $m$ the exact dispersion relation was found to take the form \cite{Babichenko:2009dk,OhlssonSax:2011ms,Sax:2012jv,Hoare:2013lja,Lloyd:2014bsa,Borsato:2014hja}
\bea
\label{eq:disp}
\varepsilon = \sqrt{m^2 + 4 h^2 \sin^2(\frac{p}{2}) }\,.
\eea
By scaling the momenta as $p/h$, together with the identification $h=g$, the large coupling expansion should agree with perturbative calculations. 

For this calculation we have to sum three distinct classes of Feynman diagrams. The first one is the sunset-type diagrams
\bea \nn
\parbox[top][0.8in][c]{1.5in}{\fmfreuse{sunset}}
\eea
giving rise to integrals of the form
\bea \nn
I^{rstu}_{m_1m_2m_3}(p) = \int \frac{d^2k d^2l}{(2\pi)^4} \frac{k_+^r k_-^s l_+^t l_-^u}{(k^2-m_1^2)(l^2-m_2^2)((p-k-l)^2-m_3^2)}\,.
\eea
The sunset contribution contains the meat of the calculation since they constitute the only class of diagrams that can give a (regularization independent) finite contribution to the amplitude. 

The other two types of diagrams are the bubble-tadpole
\bea
\nn
\parbox[top][0.8in][c]{1.5in}{\fmfreuse{doubblebubble}}
\eea
and six-vertex double tadpoles\footnote{Since we do not have the $\theta^6$-terms in the Lagrangian we can only compute these contributions for the bosons. For the fermions we will simply assume that all divergences cancel.}
\bea \nn
\parbox[top][0.8in][c]{1.5in}{\fmfreuse{doubletadpole}}
\eea 
and they are both UV-divergent and give rise to $1/\epsilon$-terms that cancel divergences coming from the sunset diagrams. 

In order to evaluate the sunset integrals we use a tensor reduction scheme similar to the one described for the one-loop bubble integrals in the previous section. We won't give the details here but rather refer to \cite{Sundin:2014ema,Sundin:2015uva}. Here we will only state the final result. 

For the massive modes we find a correction
\bea
\langle \bar z z\rangle^{(2)}=\langle \bar y y\rangle^{(2)}=\langle \bar \chi^1 \chi^1 \rangle^{(2)}=\langle \bar \chi^2 \chi^2 \rangle^{(2)}= -\frac{16}{3 g^2} p_1^4 I^{0000}_{111}(p)\vert_{p^2=1} = -\frac{p_1^4}{12 g^2}\,.
\eea
The fact that the two-loop correction is the same for bosonic and fermionic modes is a strong indication that the worldsheet theory is, as expected, supersymmetric to this loop order and that the regularization employed is consistent with this. It is easy to see that this correction precisely agrees with the exact dispersion relation (\ref{eq:disp}). 

The situation becomes more interesting in the case of massless modes. The calculation gives
\bea
\langle \bar u_1 u_1 \rangle^{(2)} = \langle \bar u_2 u_2 \rangle^{(2)}
=\langle \bar \chi^3\chi^3 \rangle^{(2)}
=\langle \bar \chi^4\chi^4 \rangle^{(2)}
 = -\frac{4}{g^2} p_1^3 I^{0100}_{011}(p)\vert_{p^2=0} = -\frac{p_1^4}{2 \pi^2 g^2}\,.
\eea
While this result is again consistent with worldsheet supersymmetry it disagrees with the expansion of the exact dispersion relation (\ref{eq:disp}). This mismatch of $6/\pi^2$ was already found for the bosonic modes in \cite{Sundin:2014ema} but here we demonstrate for the first time that the same result appears also for the fermionic modes. Although expected this is nevertheless an important consistency check.

\section{One-loop S-matrix}\label{sec:S-matrix}
Here we compute the two body S-matrix up to the one loop level. The S-matrix splits into three sectors according to whether the external particles are both massive, one massive and one massless or both massless. For each sector the contributing diagram topologies are a tree-level contact diagram, one-loop $s,t$ and $u$-channel diagrams 
\bea \label{diagram:stu}
\parbox[top][0.8in][c]{1.2in}{\fmfreuse{schannel}}+\parbox[top][0.8in][c]{1.2in}{\fmfreuse{tchannel}}
+\parbox[top][0.8in][c]{1.2in}{\fmfreuse{uchannel}}
%+\parbox[top][0.8in][c]{1.5in}{\fmfreuse{tadpolesix}}
\eea
and a six-vertex tadpole diagram
\bea
\label{diagram:t6}
\parbox[top][0.8in][c]{1.5in}{\fmfreuse{tadpolesix}}
\eea 

When calculating the S-matrix we must take into account the wave function renormalization of the massive bosons (\ref{eq:Zs}) and the Jacobian from the energy-momentum conservation delta functions and external leg factors which give a factor
%\paragraph{Agrees with \cite{Engelund:2013fja}}
\bea
J=\frac{1}{4}\frac{1}{\omega_q p- \omega_p q}\,,\qquad\omega_p=\sqrt{m^2_p+p^2}
\eea
and similarly for $\omega_q$ where $m_p$ and $m_q$ denote the masses of the corresponding external particles. We write
\begin{equation}
S=1+i\mathbbm{T}
\end{equation}
and compute the action of $\mathbbm{T}$ on generic\footnote{To save some space and time we will give only the action on either BF or FB in-states but not both. We will also omit the action on FF in-states for the most part. These additional amplitudes contain no new information and are sometimes tedious to compute.} two-particle states. 

\subsection{Massive sector}
For scattering up to one loop we find\footnote{Additional S-matrix elements can be obtained from the fact that the theory is symmetric under interchange of fields with their complex conjugates.}
\begin{align}
\textbf{Boson-Boson:}\\ \nn
\mathbbm{T} \ket{z(p) z(q)}=&\,\big(-\ell_1^{(0)}+\frac{i}{2}(\ell_1^{(0)})^2+2\theta_{LL}^{(1)} \big)\ket{z(p)z(q)}\,,\\ \nn
\mathbbm{T} \ket{y(p) y(q)}=&\,\big(\ell_1^{(0)}+\frac{i}{2}(\ell_1^{(0)})^2+2\theta_{LL}^{(1)} \big)\ket{y(p)y(q)}\,, \\ \nn
\mathbbm{T} \ket{z(p) y(q)}=&\,\big(\ell_2^{(0)}+\frac{i}{2}(\ell_2^{(0)})^2+2\theta_{LR}^{(1)} \big)\ket{z(p)y(q)}\,, \\ \nn
\mathbbm{T} \ket{y(p) z(q)}=&\,\big(-\ell_2^{(0)}+\frac{i}{2}(\ell_2^{(0)})^2+2\theta_{LR}^{(1)} \big)\ket{y(p)z(q)}\,, \\ \nn
\mathbbm{T} \ket{z(p) \bar z(q)}=&\,\big(-\ell_3^{(0)}+\frac{i}{2}[(\ell_3^{(0)})^2+2(\ell_7^{(0)})^2]+2\theta_{LR}^{(1)} \big)\ket{z(p)\bar z(q)} 
-i(\ell_7^{(0)})^2\ket{\bar y(p) y(q)}
\\ \nn
&{}+\big(\ell_7^{(0)}-\frac{i}{2}\ell_3^{(0)}\ell_7^{(0)}\big) \big[\ket{\chi^1(p) \bar \chi^1(q)}+\ket{\bar \chi^2(p) \chi^2(q)}\big]\,, \\ \nn
 \mathbbm{T} \ket{y(p) \bar y(q)}=&\,
 \big(\ell_3^{(0)}+\frac{i}{2}[(\ell_3^{(0)})^2+2(\ell_7^{(0)})^2]+2\theta_{LR}^{(1)} \big)\ket{y(p)\bar y(q)}
-i(\ell_7^{(0)})^2\ket{\bar z(p) z(q)}
\\ \nn
 &{}+\big(-\ell_7^{(0)}-\frac{i}{2}\ell_3^{(0)}\ell_7^{(0)}\big) \big[\ket{\bar\chi^1(p)\chi^1(q)}+\ket{\chi^2(p) \bar \chi^2(q)}\big]\,, \\ \nn
\mathbbm{T} \ket{z(p) \bar y(q)}=&\,
\big(\ell_2^{(0)}+\frac{i}{2}[(\ell_2^{(0)})^2+2(\ell_8^{(0)})^2]+2 \theta_{LL}^{(1)}\big)\ket{z(p) \bar y(p)}
+i(\ell_8^{(0)})^2\ket{\bar y(p) z(q)}
\\ \nn
&{}+\big(\ell_8^{(0)}+\frac{i}{2}\ell_2^{(0)}\ell_8^{(0)}\big) \big[\ket{\chi^1(p) \bar \chi^2(q)}-\ket{\bar \chi^2(p) \chi^1(q)}\big]\,, \\ \nn
\mathbbm{T} \ket{y(p) \bar z(q)}=&\,
\big(-\ell_2^{(0)}+\frac{i}{2}[(\ell_2^{(0)})^2+2(\ell_8^{(0)})^2]+2\theta_{LL}^{(1)}\big)\ket{y(p) \bar z(p)}
+i(\ell_8^{(0)})^2\ket{\bar z(p) y(q)}
\\ \nn
&{}+\big(\ell_8^{(0)}-\frac{i}{2}\ell_2^{(0)}\ell_8^{(0)}\big)\big[ \ket{\bar \chi^1(p) \chi^2(q)}-\ket{ \chi^2(p) \bar \chi^1(q)}\big]\,,\\ \nn
\textbf{Boson-Fermion:}\\ \nn
\mathbbm{T} \ket{z(p) \chi^1(q) } = &\, \big(-\ell_4^{(0)}+\frac{i}{2}[(\ell_4^{(0)})^2+(\ell_8^{(0)})^2]+2 \theta_{LL}^{(1)}\big) \ket{z(p) \chi^1(q) } \\ \nn
&{}+
\big(-\ell_8^{(0)}+\frac{i}{2}[\ell_4^{(0)}\ell_8^{(0)}+\ell_5^{(0)}\ell_8^{(0)}]\big) \ket{\chi^1(p) z(q) }, \\ \nn
\mathbbm{T} \ket{z(p) \bar \chi^1(q) } = &\, \big(\ell_6^{(0)}+\frac{i}{2}[(\ell_6^{(0)})^2+(\ell_7^{(0)})^2]+2 \theta_{LR}^{(1)}\big) \ket{z(p) \bar \chi^1(q) } \\ \nn
&{}+
\big(\ell_7^{(0)}+\frac{i}{2}[\ell_6^{(0)}\ell_7^{(0)}+\ell_7^{(0)}\ell_9^{(0)}]\big) \ket{\bar\chi^2(p) y(q) },
\\ \nn
\mathbbm{T} \ket{z(p) \chi^2(q) } = &\, \big(\ell_6^{(0)}+\frac{i}{2}[(\ell_6^{(0)})^2+(\ell_7^{(0)})^2]+2 \theta_{LR}^{(1)}\big) \ket{z(p) \chi^2(q) } \\ \nn
&{}+
\big(-\ell_7^{(0)}-\frac{i}{2}[\ell_6^{(0)}\ell_7^{(0)}+\ell_7^{(0)}\ell_9^{(0)}]\big) \ket{\chi^1(p) y(q) }, \\ \nn
\mathbbm{T} \ket{z(p) \bar \chi^2(q) } = &\, \big(-\ell_4^{(0)}+\frac{i}{2}[(\ell_4^{(0)})^2+(\ell_8^{(0)})^2]+2 \theta_{LL}^{(1)}\big) \ket{z(p) \bar \chi^2(q) } \\ \nn
&{}+
\big(-\ell_8^{(0)}+\frac{i}{2}[\ell_4^{(0)}\ell_8^{(0)}+\ell_5^{(0)}\ell_8^{(0)}]\big) \ket{\bar\chi^2(p) z(q) },
\\ \nn
\mathbbm{T} \ket{y(p) \chi^2(q) } = &\, \big(\ell_4^{(0)}+\frac{i}{2}[(\ell_4^{(0)})^2+(\ell_8^{(0)})^2]+2 \theta_{LL}^{(1)}\big) \ket{y(p) \chi^2(q) } \\ \nn
&{}+
\big(\ell_8^{(0)}+\frac{i}{2}[\ell_4^{(0)}\ell_8^{(0)}+\ell_5^{(0)}\ell_8^{(0)}]\big) \ket{\chi^2(p) y(q) }, \\ \nn
\mathbbm{T} \ket{y(p) \bar \chi^2(q) } = &\, \big(-\ell_6^{(0)}+\frac{i}{2}[(\ell_6^{(0)})^2+(\ell_7^{(0)})^2]+2 \theta_{LR}^{(1)}\big) \ket{y(p) \bar \chi^2(q) } \\ \nn
&{}+
\big(-\ell_7^{(0)}+\frac{i}{2}[\ell_6^{(0)}\ell_7^{(0)}+\ell_7^{(0)}\ell_9^{(0)}]\big) \ket{\bar\chi^1(p) z(q) },
\\ \nn
\mathbbm{T} \ket{y(p) \chi^1(q) } = &\, \big(-\ell_6^{(0)}+\frac{i}{2}[(\ell_6^{(0)})^2+(\ell_7^{(0)})^2]+2 \theta_{LR}^{(1)}\big) \ket{y(p) \chi^1(q) } \\ \nn
&{}+
\big(\ell_7^{(0)}-\frac{i}{2}[\ell_6^{(0)}\ell_7^{(0)}+\ell_7^{(0)}\ell_9^{(0)}]\big) \ket{\chi^2(p) z(q) }, \\ \nn
\mathbbm{T} \ket{y(p) \bar \chi^1(q) } = &\, \big(\ell_4^{(0)}+\frac{i}{2}[(\ell_4^{(0)})^2+(\ell_8^{(0)})^2]+2 \theta_{LL}^{(1)}\big) \ket{y(p) \bar \chi^1(q) } \\ \nn
&{}+
\big(\ell_8^{(0)}+\frac{i}{2}[\ell_4^{(0)}\ell_8^{(0)}+\ell_5^{(0)}\ell_8^{(0)}]\big) \ket{\bar\chi^1(p) y(q) },
\end{align}
where the superscript denotes tree-level or one-loop contributions and we have used (\ref{eq:massive-int}). The phase factor, $\theta$, only contributes to diagonal scattering elements. At the one-loop level, where the phase and matrix factors come with different powers of $i$, we have separated the contributions in different expressions. 

The tree-level elements are given by
\bea
&& \ell^{(0)}_1 = \frac{1}{2g} \frac{(p+q)^2}{\omega_q p - \omega_p q}\,,
\quad 
\ell_2^{(0)}=\frac{1}{2g} \frac{p^2-q^2}{\omega_q p - \omega_p q}\,,
\quad
\ell^{(0)}_3 = \frac{1}{2g} \frac{(p-q)^2}{\omega_q p - \omega_p q}\,, \\ \nn
&& \ell_4^{(0)}=\frac{1}{2g}\frac{q(p+q)}{\omega_q p -\omega_p q}\,,
\qquad \ell_5^{(0)} = \frac{1}{2g}\frac{p(p+q)}{\omega_q p- \omega_pq}\,,
\qquad \ell_6^{(0)} = \frac{1}{2g}\frac{q(p-q)}{\omega_q p- \omega_pq}
 \\ \nn
&& 
\ell_7^{(0)}=\frac{1}{2g}\frac{qp}{\omega_q p -\omega_p q}\big(\sqrt{(\omega_p-p)(\omega_q+q)}-\sqrt{(\omega_p+p)(\omega_q-q)}\big)\,, \\ \nn
&& \ell_8^{(0)}=\frac{1}{2g}\frac{qp}{\omega_q p -\omega_p q}\big(\sqrt{(\omega_p-p)(\omega_q+q)}+\sqrt{(\omega_p+p)(\omega_q-q)}\big)\,, 
\qquad \ell_9^{(0)} = \frac{1}{2g}\frac{p(p-q)}{\omega_q p- \omega_pq}
\eea
where $\ell_5^{(0)}$ and $\ell_9^{(0)}$ appear at tree-level only in FB processes which we have not listed. The two one-loop phases are given by
\bea
\label{eq:massive-phases}
&& \theta_{LL}^{(1)} = -\frac{1}{4\pi}\frac{1}{g^2}\left(\frac{p^2q^2\big(\textbf{p}\cdot \textbf{q} +1\big)}{(\omega_q p - \omega_p q)^2}\log\frac{q_-}{p_-}+\frac12\frac{p q(p+q)^2}{\omega_q p - \omega_p q}\right)\,, \\ \nn
&& \theta_{LR}^{(1)} = -\frac{1}{4\pi}\frac{1}{g^2}\left(\frac{p^2q^2\big(\textbf{p}\cdot \textbf{q}-1 \big)}{(\omega_q p - \omega_p q)^2}\log\frac{q_-}{p_-}-\frac12\frac{p q(p-q)^2}{\omega_p q - \omega_q p}\right)\,.
\eea
While we do not present the details here, it is straightforward to compare and see that our findings fit perfectly with the exact S-matrix proposed in \cite{Borsato:2013qpa} and phases proposed in \cite{Borsato:2013hoa} (this was mostly checked already in \cite{Roiban:2014cia}).\footnote{Their excitations are related to ours as follows:
$$
Z^R=z\,,\quad Z^L=\bar z\,,\quad Y^R=\bar y\,,\quad Y^L=y\,,\qquad\eta^R_1=\chi^1\,,\quad\eta^R_2=\bar\chi^2\,,\quad\eta^{1L}=\bar\chi^1\,,\quad\eta^{2L}=\chi^2\,.
$$
}

\subsection{Mixed sector}
Next we turn to the sector where we scatter one massive and one massless particle. For simplicity we will restrict to in-states where the first particle is massive and the second massless. When the masses of the scattered particles are different, the energy-momentum constraints allow also for a more complicated solution where the momenta of the particles in the final state differ from those in the initial state. However, only the trivial solution where each particles momentum is unchanged is compatible with higher conserved charges and integrability. This means that if we were to find any non-zero on-shell amplitudes where the particles have changed their momenta integrability would be broken at one loop. 

To be a bit more explicit, let us call the non-trivial solutions of the energy-momentum constraints $\tilde p(p,q)$ and $\tilde q(p,q)$, where as before $p$ and $q$ are the in-coming momenta. We furthermore introduce the notation $X_i$ and $x_i$ for arbitrary $i,j$-flavored massive / massless particles. The statement above then implies that if we find a non-zero scattering element, 
\bea \nn
\mathbbm{T} \ket {X_i(p) x_j(q) } \rightarrow A_{ij}^{kl}(p,q) \ket{X_k (\tilde p) x_l (\tilde q)}\,,
\eea
with $\tilde p\neq p$ then integrability is broken. We have verified numerically for a large class of scattering processes that $A_{ij}^{kl}(p,q)=0$, consistent with integrability at the one-loop level.

The non-zero amplitudes we find, in the kinematic regime $(p>0>q)$, are:
\begin{align}
\textbf{Boson-Boson} \\ \nn
\mathbbm{T} \ket{z(p)  u_1(q) }=&\, \big(\ell_1^{(0)}+\frac{i}{2}[(\ell_1^{(0)})^2+(\ell_2^{(0)})^2]+\theta^{(1)}\big) \ket{z(p) u_1(q)}\\ \nn
 &\,{}+\big(-\ell_2^{(0)}+\frac{i}{2}[\ell_2^{(0)}\ell_4^{(0)}-2\ell_2^{(0)}\ell_6^{(0)}]\big) \ket{\bar\chi^2(p) \bar \chi^4(q)}\,, \\ \nn
\mathbbm{T} \ket{z(p)  \bar u_1(q) } =&\,\big(\ell_1^{(0)}+\frac{i}{2}[(\ell_1^{(0)})^2+(\ell_2^{(0)})^2]+\theta^{(1)}\big) \ket{z(p) \bar u_1(q)}\\ \nn
&\,{}+\big(\ell_2^{(0)}-\frac{i}{2}[\ell_2^{(0)}\ell_4^{(0)}-2\ell_2^{(0)}\ell_6^{(0)}]\big)\ket{\chi^1(p) \bar \chi^3(q)}\,, \\ \nn
\mathbbm{T} \ket{z(p)  u_2(q) } =&\,\big(\ell_1^{(0)}+\frac{i}{2}[(\ell_1^{(0)})^2+(\ell_2^{(0)})^2]+\theta^{(1)}\big) \ket{z(p) u_2(q)}\\ \nn
 &\,{}+\big(\ell_2^{(0)}-\frac{i}{2}[\ell_2^{(0)}\ell_4^{(0)}-2\ell_2^{(0)}\ell_6^{(0)}]\big) \ket{ \bar \chi^2(p) \bar \chi^3(q)}\,,\\ \nn
 \mathbbm{T} \ket{z(p)  \bar u_2(q) } =&\, \big(\ell_1^{(0)}+\frac{i}{2}[(\ell_1^{(0)})^2+(\ell_2^{(0)})^2]+\theta^{(1)}\big) \ket{z(p) \bar u_2(q)} \\ \nn
&\,{}+\big(\ell_2^{(0)}-\frac{i}{2}[\ell_2^{(0)}\ell_4^{(0)}-2\ell_2^{(0)}\ell_6^{(0)}]\big) \ket{ \chi^1(p) \bar \chi^4(q)}\,, \\ \nn
\mathbbm{T} \ket{y(p)  u_1(q) } =&\, \big(-\ell_1^{(0)}+\frac{i}{2}[(\ell_1^{(0)})^2+(\ell_2^{(0)})^2]+\theta^{(1)}\big) \ket{y(p) u_1(q)}\\ \nn
&\,{}+\big(\ell_2^{(0)}+\frac{i}{2}[\ell_2^{(0)}\ell_4^{(0)}-2\ell_2^{(0)}\ell_6^{(0)}]\big) \ket{\bar\chi^1(p) \bar \chi^4(q)}\,, \\ \nn
\mathbbm{T} \ket{y(p)  \bar u_1(q) } =&\, \big(-\ell_1^{(0)}+\frac{i}{2}[(\ell_1^{(0)})^2+(\ell_2^{(0)})^2]+\theta^{(1)}\big) \ket{y(p) \bar u_1(q)}\\ \nn
&\,{}+\big(\ell_2^{(0)}+\frac{i}{2}[\ell_2^{(0)}\ell_4^{(0)}-2\ell_2^{(0)}\ell_6^{(0)}]\big)\ket{\chi^2(p) \bar \chi^3(q)}\,, \\ \nn
\mathbbm{T} \ket{y(p)  u_2(q) } =&\,\big(-\ell_1^{(0)}+\frac{i}{2}[(\ell_1^{(0)})^2+(\ell_2^{(0)})^2]+\theta^{(1)}\big) \ket{y(p) u_2(q)}\\ \nn
&\,{}+\big(-\ell_2^{(0)}-\frac{i}{2}[\ell_2^{(0)}\ell_4^{(0)}-2\ell_2^{(0)}\ell_6^{(0)}]\big) \ket{ \bar \chi^1(p) \bar \chi^3(q)}\,, \\ \nn
 \mathbbm{T} \ket{y(p)  \bar u_2(q) } =&\, \big(-\ell_1^{(0)}+\frac{i}{2}[(\ell_1^{(0)})^2+(\ell_2^{(0)})^2]+\theta^{(1)}\big) \ket{y(p) \bar u_2(q)}\\ \nn
&\,{}+\big(\ell_2^{(0)}+\frac{i}{2}[\ell_2^{(0)}\ell_4^{(0)}-2\ell_2^{(0)}\ell_6^{(0)}]\big) \ket{ \chi^2(p) \bar \chi^4(q)}\,, \\ \nn
\textbf{Boson-Fermion} \\ \nn
 \mathbbm{T} \ket{z(p)\chi^3(q)}=&\,\big(-\ell_4^{(0)}+\frac{i}{2}[(\ell_4^{(0)})^2+2(\ell_2^{(0)})^2]+\theta^{(1)}\big) \ket{z(p)\chi^3(q)}\\ \nn
&\,{}+(\ell_2^{(0)}-\frac{i}{2}\ell_2^{(0)}\ell_4^{(0)})\big(\ket{\chi^1(p) u_1(q)}+\ket{\bar \chi^2(p) \bar u_2(q)}\big)+i(\ell_2^{(0)})^2\ket{\bar y(p) \bar \chi^4(q)}\,, \\ \nn
\mathbbm{T}\ket{z(p) \bar \chi^{3,4}(q)}=&\,\big(\ell_3^{(0)}+\frac{i}{2}(\ell_3^{(0)})^2+\theta^{(1)}\big) \ket{z(p) \bar \chi^{3,4}(q)}\,, \\ \nn
\mathbbm{T} \ket{z(p)\chi^4(q)} =&\,
 \big(-\ell_4^{(0)}+\frac{i}{2}[(\ell_4^{(0)})^2+2(\ell_2^{(0)})^2]+\theta^{(1)}\big) \ket{z(p)\chi^4(q)}\\ \nn
&\,{}+(\ell_2^{(0)}-\frac{i}{2}\ell_2^{(0)}\ell_4^{(0)})\big(\ket{\chi^1(p) u_2(q)}-\ket{\bar \chi^2(p) \bar u_1(q)}\big)-i(\ell_2^{(0)})^2\ket{\bar y(p) \bar \chi^3(q)}\,, \\ \nn
%%%%%%
\mathbbm{T} \ket{y(p)\chi^3(q)} =&\,\big(\ell_4^{(0)}+\frac{i}{2}[(\ell_4^{(0)})^2+2(\ell_2^{(0)})^2]+\theta^{(1)}\big) \ket{y(p)\chi^3(q)}\\ \nn
&\,{}+ (\ell_2^{(0)}+\frac{i}{2}\ell_2^{(0)}\ell_4^{(0)})\big(\ket{\chi^2(p) u_1(q)}-\ket{\bar \chi^1(p) \bar u_2(q)}\big)-i(\ell_2^{(0)})^2\ket{\bar z(p) \bar \chi^4(q)}\,, \\ \nn
\mathbbm{T}\ket{y(p) \bar \chi^{3,4}(q)} =&\,\big(-\ell_3^{(0)}+\frac{i}{2}(\ell_3^{(0)})^2+\theta^{(1)}\big) \ket{y(p) \bar \chi^{3,4}(q)} \\ \nn
\mathbbm{T} \ket{y(p)\chi^4(q)} =&\, 
\big(\ell_4^{(0)}+\frac{i}{2}[(\ell_4^{(0)})^2+2(\ell_2^{(0)})^2]+\theta^{(1)}\big) \ket{y(p)\chi^4(q)}\\ \nn
&\,{}+(\ell_2^{(0)}+\frac{i}{2}\ell_2^{(0)}\ell_4^{(0)})\big(\ket{\chi^2(p) u_2(q)}+\ket{\bar \chi^1(p) \bar u_1(q)}\big)+i(\ell_2^{(0)})^2\ket{\bar z(p) \bar \chi^3(q)}\,, \\ \nn
\mathbbm{T} \ket{\chi^1(p) u_1(q)} =&\, 
\big(i(\ell_2^{(0)})^2+\theta^{(1)}\big) \ket{\chi^1(p) u_1(q)}+i(\ell_2^{(0)})^2\ket{\bar \chi^2(p) \bar u_2(q)} \\ \nn
&\,{}+ \big(\ell_2^{(0)}-\frac{i}{2} \ell_2^{(0)}\ell_4^{(0)}\big) \ket{z(p) \chi^3(q)} + \big(\ell_2^{(0)}+\frac{i}{2}\ell_2^{(0)}\ell_4^{(0)}\big) \ket{\bar y(p) \bar \chi^4(q)}\,, \\ \nn
\mathbbm{T} \ket{\chi^1(p) u_2(q)} =&\,
\big(i(\ell_2^{(0)})^2+\theta^{(1)}\big) \ket{\chi^1(p) u_2(q)}-i(\ell_2^{(0)})^2\ket{\bar \chi^2(p) \bar u_1(q)} \\ \nn
&\,{}+ \big(\ell_2^{(0)}-\frac{i}{2} \ell_2^{(0)}\ell_4^{(0)}\big) \ket{z(p) \chi^4(q)} + \big(-\ell_2^{(0)}-\frac{i}{2} \ell_2^{(0)}\ell_4^{(0)}\big) \ket{\bar y(p) \bar \chi^3(q)}\,, \\ \nn
\mathbbm{T} \ket{\chi^2(p) u_1(q)} =&\, 
\big(i(\ell_2^{(0)})^2+\theta^{(1)}\big) \ket{\chi^2(p) u_1(q)}-i(\ell_2^{(0)})^2\ket{\bar \chi^1(p) \bar u_2(q)} \\ \nn
&\,{}+ \big(\ell_2^{(0)}+\frac{i}{2} \ell_2^{(0)}\ell_4^{(0)}\big) \ket{y(p) \chi^3(q)} + \big(-\ell_2^{(0)}+\frac{i}{2} \ell_2^{(0)}\ell_4^{(0)}\big) \ket{\bar z(p) \bar \chi^4(q)}\,, \\ \nn
\mathbbm{T} \ket{\chi^2(p) u_2(q)} =&\,
\big(i(\ell_2^{(0)})^2+\theta^{(1)}\big) \ket{\chi^2(p) u_2(q)}+i(\ell_2^{(0)})^2\ket{\bar \chi^1(p) \bar u_1(q)} \\ \nn
&\,{}+ \big(\ell_2^{(0)}+\frac{i}{2}\ell_2^{(0)} \ell_4^{(0)}\big) \ket{y(p) \chi^4(q)} + \big(\ell_2^{(0)}-\frac{i}{2} \ell_2^{(0)}\ell_4^{(0)}\big) \ket{\bar z(p) \bar \chi^3(q)}\,, \\ \nn
\mathbbm{T} \ket{\chi^1(p) \bar u_1(q)}=&\,\theta^{(1)}\ket{\chi^1(p) \bar u_1 (q) }\,,\\ \nn
\mathbbm{T} \ket{\chi^2(p) \bar u_2(q)}=&\,\theta^{(1)}\ket{\chi^2(p) \bar u_2 (q) }\,,
\end{align}
where we have used (\ref{eq:mixing-int-t}) and (\ref{eq:mixing-int-u}). The tree-level amplitudes involve
\bea
&&\ell_1^{(0)}=\frac{1}{2g}\frac{q}{p+\omega_p}\,, \qquad
\ell_2^{(0)}=\frac{1}{2g}p \sqrt{2(p-\omega_p)q}\,,   \qquad
\ell_3^{(0)}=\frac{1}{2g}\frac{p+q}{p+\omega_p}\,, \\ \nn
 &&
\ell_4^{(0)}=\frac{1}{2g}\frac{p-q}{p+\omega_p}\,, \qquad 
\ell_6^{(0)}=\frac{1}{2g}\frac{p}{p+\omega_p}\,,
\eea
where $\ell^{(0)}_6$ is related to purely fermionic tree-level scattering. The one-loop phase in the mixed sector is given by
\begin{equation}
\theta^{(1)}=-\frac{1}{2\pi}\frac{1}{g^2} \frac{(1-\log \frac{q_-}{p_-})p^2 q}{p+\omega_p}\,.
\label{eq:mixed-phase}
\end{equation}
Ignoring for the moment the one-loop phase, which will be discussed in the next section, the S-matrix in the mixed sector does not agree with the exact S-matrix proposed in \cite{Borsato:2014hja} if one assumes their dressing phase is trivial at tree-level.\footnote{Their massless excitations are related to ours as follows:
$$
T^{11}=u_2\,,\quad T^{12}=u_1\,,\quad T^{21}=-\bar u_1\,,\quad T^{22}=\bar u_2\,,
$$
$$
\chi^1(p)=-\mathrm{sign}(p)\bar\chi^3(p)\,,\quad\chi^2=-\bar\chi^4\,,\quad\tilde\chi^1(p)=-\mathrm{sign}(p)\chi^4(p)\,,\quad\tilde\chi^2=\chi^3\,.
$$
Note the sign changes between left and right-moving massless fermions.
\label{foot:massless}
}
 However, this disagreement can be removed by making the dressing phase non-trivial at tree-level or, equivalently, changing the normalization of their S-matrix, which is not fixed by the symmetries, by a (crossing invariant) rational phase\footnote{We thank Olof Ohlsson Sax for discussions of this.}
\begin{equation}
\frac{x^-}{x^+}
\frac{y^+}{y^-}
\frac{x^+-y^-}{x^--y^+}
\frac{1-\frac{1}{x^-y^+}}{1-\frac{1}{x^+y^-}}\,.
\end{equation}

\subsection{Massless sector}
We now turn to the massless sector. Since the S-matrix is defined with respect to asymptotic states, determined by the leading order relativistic Lagrangian, we will assume that the transverse momenta satisfy $p>0>q$ throughout this section. 

Summing all the diagrams one finds
\begin{align}
\textbf{Boson-Boson}\\ \nn
\mathbbm{T} \ket{u_i(p) u_j(q)} =&\,\theta^{(1)} \ket{u_i(p)u_j(q)}\,, \\ \nn
\mathbbm{T} \ket{u_1(p)\bar u_1(q)} =&\,\big(\theta^{(1)} +i(\ell^{(0)})^2\big)\ket{u_1(p)\bar u_1(q)} + i(\ell^{(0)})^2 \ket{\bar u_2(p)u_2(q)} -\ell^{(0)}  \ket{\chi^3(p) \bar \chi^3(q)}\\ \nn
&\,{}+\ell^{(0)} \ket{\bar \chi^4(p) \chi^4(q)}\,, \\ \nn
\mathbbm{T} \ket{u_2(p)\bar u_2(q)} =&\,\big(\theta^{(1)} +i(\ell^{(0)})^2\big) \ket{u_2(p)\bar u_2(q)} + i(\ell^{(0)})^2 \ket{\bar u_1(p)u_1(q)} +\ell^{(0)} \ket{\chi^4(p) \bar \chi^4(q)}\\ \nn
&\,{}-\ell^{(0)} \ket{\bar \chi^3(p) \chi^3(q)}\,,\\ \nn 
\mathbbm{T} \ket{u_1(p)\bar u_2(q)} =&\,\big(\theta^{(1)} +i(\ell^{(0)})^2\big) \ket{u_1(p)\bar u_2(q)} -i(\ell^{(0)})^2 \ket{\bar u_2(p)u_1(q)} -\ell^{(0)} \ket{\chi^3(p)\bar \chi^4(q)}\\ \nn
&\,{}-\ell^{(0)} \ket{\bar \chi^4(p) \chi^3(q)}\,, \\ \nn
\mathbbm{T} \ket{u_2(p)\bar u_1(q)} =&\,\big(\theta^{(1)} +i(\ell^{(0)})^2\big) \ket{u_2(p)\bar u_1(q)} -i(\ell^{(0)})^2 \ket{\bar u_1(p)u_2(q)}+\ell^{(0)} \ket{\chi^4(p)\bar \chi^3(q)}\\ \nn
&\,{}+\ell^{(0)} \ket{\bar \chi^3(p) \chi^4(q)}, \\ \nn
\textbf{Boson-Fermion} \\ \nn
%&&
%\mathbbm{T}\ket{\bar u_i(p) \chi^k(q)} = \big(i\theta^{(1)} + \frac{1}{2} (i\ell^{(0)})^2\big)\ket{\bar u_i(p) \chi^k(q)}- i\ell^{(0)} \big(\delta^k_4 \delta_i^1-\delta^k_3 \delta_i^2\big)\ket{ \chi^k(p) \bar u_i(q)}, \\ \nn
\mathbbm{T}\ket{u_1(p) \chi^3(q)} =&\,\big(\theta^{(1)} + \frac{i}{2}(\ell^{(0)})^2\big)\ket{u_1(p) \chi^3(q)}-\ell^{(0)}\ket{\chi^3(p) u_1(q)}, \\ \nn
\mathbbm{T}\ket{u_1(p) \chi^4(q)} =&\,\big(\theta^{(1)} + \frac{i}{2}(\ell^{(0)})^2\big)\ket{u_1(p) \chi^4(q)}-\ell^{(0)}\ket{\chi^3(p) u_2(q)}, \\ \nn
\mathbbm{T}\ket{u_2(p) \chi^3(q)} =&\,\big(\theta^{(1)} + \frac{i}{2}(\ell^{(0)})^2\big)\ket{u_2(p) \chi^3(q)}+\ell^{(0)}\ket{\chi^4(p) u_1(q)}, \\ \nn
\mathbbm{T}\ket{u_2(p) \chi^4(q)} =&\,\big(\theta^{(1)} + \frac{i}{2}(\ell^{(0)})^2\big)\ket{u_2(p) \chi^4(q)}+\ell^{(0)}\ket{\chi^4(p) u_2(q)}, \\ \nn
\mathbbm{T}\ket{u_1(p) \bar \chi^3(q)} =&\,\big(\theta^{(1)} + \frac{i}{2}(\ell^{(0)})^2\big)\ket{u_1(p) \bar\chi^3(q)}-\ell^{(0)}\ket{\bar \chi^4(p) u_2(q)}, \\ \nn
\mathbbm{T}\ket{u_1(p) \bar \chi^4(q)} =&\,\big(\theta^{(1)} + \frac{i}{2}(\ell^{(0)})^2\big)\ket{u_1(p) \bar\chi^4(q)}+\ell^{(0)}\ket{\bar \chi^4(p) u_1(q)}, \\ \nn
\mathbbm{T}\ket{u_2(p) \bar \chi^3(q)} =&\,\big(\theta^{(1)} + \frac{i}{2}(\ell^{(0)})^2\big)\ket{u_2(p) \bar\chi^3(q)}-\ell^{(0)}\ket{\bar \chi^3(p) u_2(q)}, \\ \nn
\mathbbm{T}\ket{u_2(p) \bar \chi^4(q)} =&\,\big(\theta^{(1)} + \frac{i}{2}(\ell^{(0)})^2\big)\ket{u_2(p) \bar\chi^4(q)}+\ell^{(0)}\ket{\bar \chi^3(p) u_1(q)}, \\ \nn
\textbf{Fermion-Fermion} \\ \nn
\mathbbm{T} \ket{\chi^i(p) \chi^j(q)} =&\,\theta^{(1)} \ket{\chi^i(p) \chi^j(q)}, \\ \nn
\mathbbm{T} \ket{\chi^i(p) \bar \chi^j(q)} =&\,\big(\theta^{(1)}+i(\ell^{(0)})^2\big) \ket{\chi^i(p) \bar \chi^j(q)}+ \dots
\end{align}
where $i,j=3,4$. Note that for fermionic four-point functions we have ignored potential divergent tadpole diagrams. In order to check that the amplitude is finite we would need the $\theta^6$ terms of the Lagrangian (which we do not have). Furthermore, for fermion in-states we have only computed the diagonal element (due to the complexity of the computation). 

The tree-level amplitudes involve
\begin{equation}
\ell^{(0)} =\frac{1}{2g} \sqrt{-pq}\,,
\end{equation}
while the one-loop phase in the massless sector is given by
\bea 
\theta^{(1)} = -\frac{1}{4\pi} \frac{1}{g^2}\big(1- \log (-4 p q)\big) p q\,.
\label{eq:massless-phase}
\eea
Ignoring for the moment the one-loop phase, which will be discussed in the next section, the structure of the S-matrix in the massless sector matches precisely with that proposed in \cite{Borsato:2014hja} (see footnote \ref{foot:massless} for the identification of excitations).

\subsection*{Two-loop boson-boson scattering}
In fact, we can push the analysis to the two-loop level for purely bosonic processes,
\bea
\ket{B_1(p) B_2(q)} \rightarrow \ket{B_3(p)B_4(q)}\,,
\eea
which feel the presence of the extra $\mathfrak{su}(2)$-factor of \cite{Borsato:2014hja}. In \cite{Sundin:2015uva} one specific element, $u_1 u_1 \rightarrow u_1 u_1$ was computed in the type IIA setting. Here we generalize the corresponding computation for type IIB to include all external massless bosons. The computation is rather complicated and involves the four-vertex wineglass diagrams
\begin{center}
\includegraphics[scale=0.6]{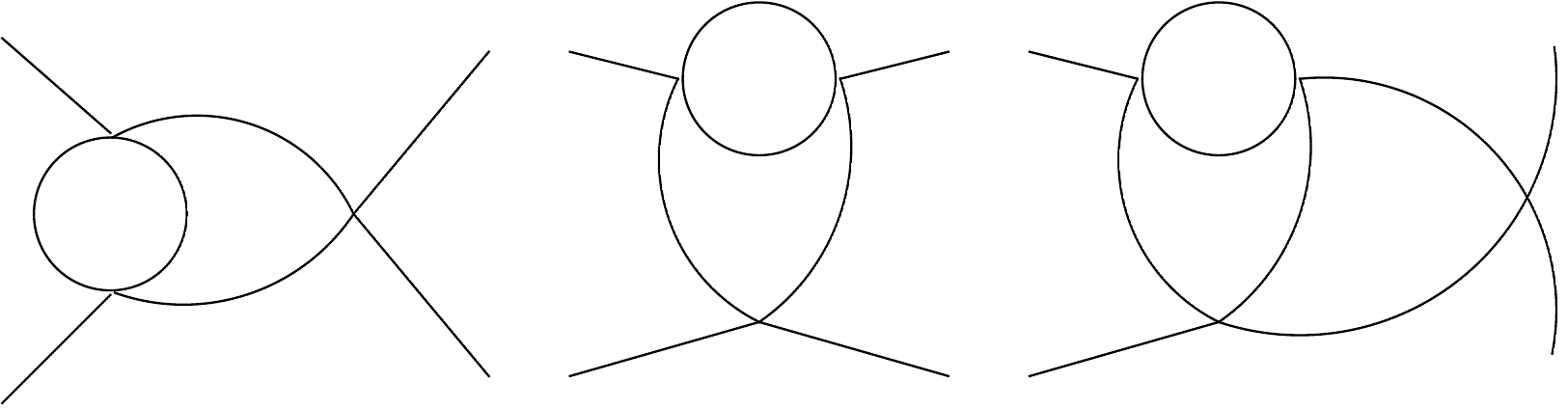}
\end{center}
plus permutations. We also have double bubbles
\begin{center}
\includegraphics[scale=0.6]{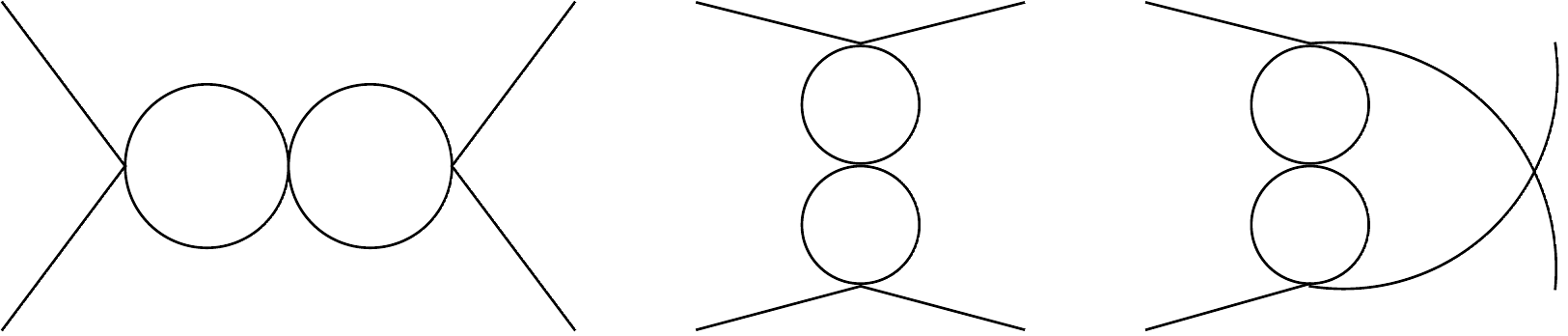}
\end{center}
and finally bubble tadpole type diagrams
\begin{center}
\includegraphics[scale=0.6]{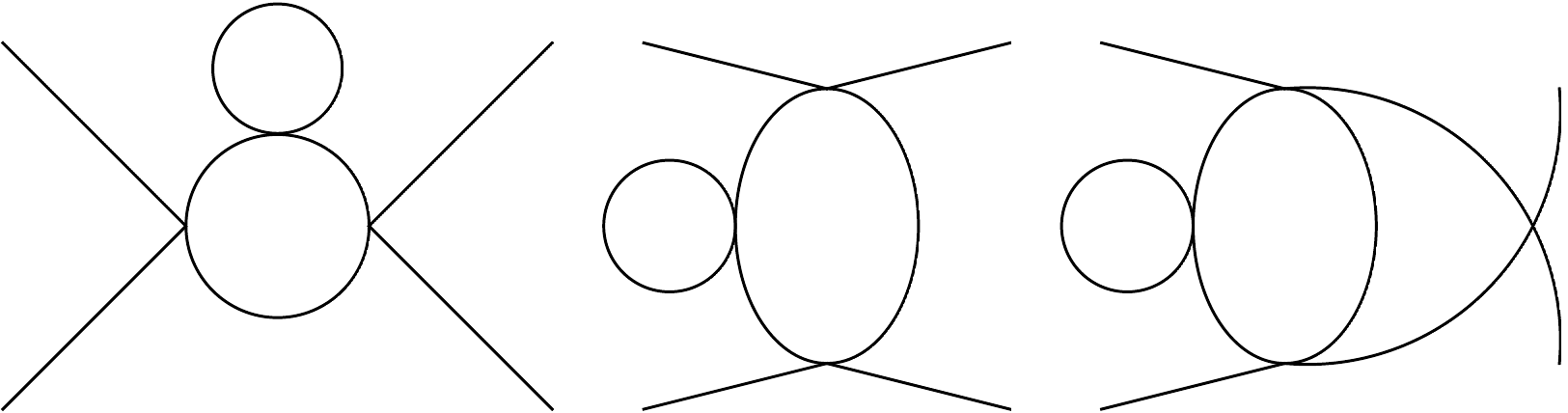}
\end{center}
There are additional diagrams involving six-vertices. The sole purpose of these types of diagrams is to render the amplitude finite. For simplicity we will ignore them here, but for the forward scattering element investigated in \cite{Sundin:2015uva} it was demonstrated that the full amplitude was indeed finite. 

It turns out that each contributing integral from the diagrams listed above comes multiplied with high enough powers of $p^2$ and $q^2$ to render everything zero once we go on-shell by setting $p^2=q^2=0$. We therefore conclude that\footnote{Note that we ignore out-states involving two massless fermions on the RHS.}
\bea
\mathbbm{T}^{(2)} \ket{u_i (p) u_j(q)}=0, \qquad
\mathbbm{T}^{(2)} \ket{u_i (p) \bar u_j(q)}=0\,.
\eea
This is compatible with the S-matrix of \cite{Borsato:2014hja} but it also implies that the total phase should vanish at two loops.
%A consequence of this is that either the two-loop zero is manifestly zero or it cancels against matrix elements. The latter alternative is not very natural tho since the phase, at two loops, both consist of a real and imaginary contribution. The matrix elements on the other hand are completely determined by the optical theorem and are purely real. 
For a more detailed discussion for the massive example of $AdS_5 \times S^5$ we point the interested reader to \cite{Klose:2007rz}.

\section{Phases and crossing symmetry}\label{sec:phases}
The crossing equation for the mixed sector phase is (for massive-massless) given by \cite{Borsato:2014hja}
\bea
(\sigma_{\bar p q})^2(\sigma_{pq})^2=\frac{x^+_p}{x^-_p}\frac{x^-_p-x^+_q}{x^+_p-x^+_q}\frac{1-\frac{1}{x^+_px^+_q}}{1-\frac{1}{x^-_px^+_q}}\,,
\eea
where $\bar p =(-\omega_p,-p)$. Expanding this expression we find
\bea
(\sigma_{\bar p q})^2(\sigma_{pq})^2=1-\frac{1}{2h^2}\frac{p^2q}{p+\omega_p}+\mathcal{O}(h^{-4})\,.
\eea
It is an easy exercise to verify that the phase found from the worldsheet computation $\sigma^2=e^{-i\theta^{(1)}}$, with $\theta^{(1)}$ given in (\ref{eq:mixed-phase}), satisfies this equation. %Note that the crossing equations are independent of the $\log$-piece of the phase. In particular this means that we can rescale the argument with any $p$ independent factor we want without changing the solution of the equations. This will be become important below. 

Next we turn to the crossing equation in the massless sector. It reads
\bea
(\sigma_{\bar p q})^2 (\sigma_{pq})^2=\frac{\varsigma_{pq}-1}{\varsigma_{pq}}\frac{1-\frac{1}{x^+_p x^+_q}}{1-\frac{1}{x^+_px^-_q}}\frac{1-\frac{1}{x^-_p x^-_q}}{1-\frac{1}{x^-_px^+_q}}\,,
\eea
where $\varsigma_{pq}$ is a scalar function of the rapidities not determined by crossing or the underlying symmetries. Expanding for large $h$ gives
\bea
%\frac{\varsigma_{pq}-1}{\varsigma_{pq}}\frac{1-\frac{1}{x^+_p x^+_q}}{1-\frac{1}{x^+_px^-_q}}\frac{1-\frac{1}{x^-_p x^-_q}}{1-\frac{1}{x^-_px^+_q}}
(\sigma_{\bar p q})^2 (\sigma_{pq})^2
=\frac{\varsigma_{pq}-1}{\varsigma_{pq}}\big(1-\frac{pq}{4h^2}\big)+\mathcal{O}(h^{-4})\,.
\eea
This is satisfied by the one-loop massless phase $\sigma^2=e^{-i\theta^{(1)}}$, with $\theta^{(1)}$ given in (\ref{eq:massless-phase}), only if the pre-factor $(\varsigma_{pq}-1)/\varsigma_{pq}$ equals $1+\mathcal O(h^{-3})$. A natural way to make this happen is to take $\varsigma_{pq}=\infty$. The crossing equation is clearly also consistent with a vanishing two-loop phase.

This choice of $\varsigma_{pq}$ is further motivated by the worldsheet scattering processes we have computed. This function enters the $\mathfrak{su}(2)$ S-matrix that acts on the massless modes and is given by \cite{Borsato:2014hja}
\bea
S_{\mathfrak{su}(2)}=\frac{1}{\varsigma_{pq}+1}\big( \varsigma_{pq}\mathbbm{1} + \Pi\big)\,,
\eea
where $\Pi$ denotes the permutation operator. However, since all one-loop elements, modulo the phase, are reproduced by the $\mathfrak{psu}(1|1)^4$ factors, the $\Pi$ operator must drop out. Furthermore, since the two-loop S-matrix vanished for purely bosonic in- and out-states we can conclude that the $\mathfrak{su}(2)$-factor must be trivial to order $h^{-3}$ in perturbation theory. As before, the most natural way to make this happen is to assume that $\varsigma=\infty$.

We will now demonstrate that the phase we found on the worldsheet can, up to an IR-divergent term, be obtained from the massive HL-phase. Introducing general masses $m_p$ and $m_q$ it is given by (this comes from the expression of the phase as a sum of charges)
\bea
\label{eq:HL-phase}
\theta_{HL}=
-\frac{1}{2\pi}\frac{1}{g^2}\frac{p^2q^2 \textbf{p}\cdot \textbf{q}}{(\omega_q p - \omega_p q)^2}\log\frac{m_p q_-}{m_q p_-}+\frac{1}{2\pi}\frac{1}{g^2}\frac{p^2 q^2}{\omega_q p - \omega_p q}\,,
\eea
where $\omega_p=\sqrt{m_p^2+p^2}$ and similarly for $\omega_q$. For $m_p=m_q=1$ this phase is identical to the sum of the LL and LR phases of the massive sector (\ref{eq:massive-phases}), $\theta_{HL} = \theta_{LL} + \theta_{LR}$. 

However, taking the appropriate massless limit to reach the mixed and massless sectors we find
\bea
\label{eq:HL-limits}
&& \textrm{Mixed sector:}\qquad \theta_{HL}(m_p=1,m_q=\mu) = -\frac{1}{2\pi}\frac{1}{g^2} \frac{(1-\log \frac{q_-}{\mu p_-})p^2 q}{p+\omega_p}\,, \\ \nn
&& \textrm{Massless sector:}\qquad \theta_{HL}(m_p=m_q=\mu) =-\frac{1}{4\pi} \frac{1}{g^2}\big(1- \log (-\frac{4 p q}{\mu^2})\big) p q\,.
\eea
Up to an IR-divergent term as $\mu\rightarrow0$, the massless limit of the HL phase thus completely reproduces the mixed and massless sector phases (\ref{eq:mixed-phase}) and (\ref{eq:massless-phase}). 

The reason that the mixed and massless sector phases arise as a limit of (the sum of) the massive phases is that the contributing integrals are just massless limits of the massive ones, see appendix \ref{sec:appendix-integrals}. The IR-divergences that appear in the cases with massless modes are canceled by tadpole contributions. To see in detail how this happens we can look at a specific scattering element, $u_1 u_1 \rightarrow  u_1 u_1$, in the massless sector. Since the IR-divergent term comes from the log-piece, it is enough to consider the $s$ and $u$-channels (the six-vertex tadpoles are all IR-finite). The contributing integrals are
\bea
\label{eq:u1-ex}
p q \left[ B^{11}_{\mu\mu}(p-q) -2 p q B^{00}_{\mu\mu}(p-q)\right] + \dots 
\eea
where we have put the external momenta on-shell. Rewriting the UV-divergent bubble integral as $B^{11}_{\mu\mu} = \mu^2 B^{00}_{\mu\mu} + T^{00}_\mu$ it's easy to see, using appendix \ref{sec:appendix-integrals}, that the above combination is IR-finite. 

However, this analysis also shows that the numerical factor inside the log in the mixed and massless sector phases is mildly regularization dependent. In principle we could use the IR-regulator $\mu$ for bubble integrals and $b\mu$, where $b$ is a constant, for tadpole integrals in which case we would get instead
\bea
\frac{1}{4\pi}\log (-4b^2 pq)pq+\dots
\eea
for the massless sector phase and one factor of $b$ in the log in the mixed sector phase. Although this regularization scheme does not seem natural from the worldsheet perspective it appears to be important in comparing to the minimal all-loop phases proposed in \cite{Borsato:2016kbm}.

\section{Concluding remarks}
In this paper we have computed the full worldsheet two-particle S-matrix for the type IIB BMN superstring in $AdS_3 \times S^3 \times T^4$ at one loop. By using integrability and the underlying symmetries of the model the full S-matrix, modulo phase factors, can be determined \cite{Borsato:2014hja} and our findings are in complete agreement with these results. We have also demonstrated that the phase factors in the mixed and massless sectors satisfy the crossing equations of \cite{Borsato:2014hja, Borsato:2016kbm} and can in a sense be seen as limits of the massive one-loop HL-phase. Furthermore, in the massless sector of the theory there is an additional $\mathfrak{su}(2)$-factor in the S-matrix of \cite{Borsato:2014hja} which we find to act trivially.

We have also pushed the analysis to the two-loop order. The symmetries of the model can be used to write down an exact form of the dispersion relation for the fundamental excitations in terms of the central charges \cite{Borsato:2014exa}. The latter have been computed at the classical level. The corresponding exact dispersion relation is expected to coincide with the pole of two-point functions of the string modes. In this paper we have extended the analysis of \cite{Sundin:2014ema, Sundin:2015uva} by including the two-point function of the fermionic modes using the type IIB string. Again we have found that the massive sector agrees with the exact proposal while there is the same mysterious disagreement in the massless sector as was found earlier.

Restricting to purely bosonic external particles we have also computed the two-loop massless S-matrix and find that it is trivial, in agreement with a similar calculation for the type IIA string where a single scattering processes was considered in \cite{Sundin:2015uva}. 

There are several natural extensions of this work. Most important is of course to understand the reason for the apparent disagreement between perturbative calculations and symmetry arguments for the two-loop dispersion relation of massless bosons and fermions.

Another interesting line of research is to consider the more general background of $AdS_3\times S^3\times S^3 \times S^1$. Many novel features such as several distinct masses and a non-trivial interpolating function $h(g)$ appears in this model \cite{Babichenko:2009dk, Sundin:2012gc,Abbott:2012dd,Beccaria:2012kb,Borsato:2012ss,Abbott:2013mpa,Borsato:2015mma}. While the integrability story is similar to the $T^4$ case, the perturbative analysis is significantly more involved since the string Lagrangian has interaction terms with an odd number of fields as well \cite{Rughoonauth:2012qd, Sundin:2012gc}.

\section*{Acknowledgements}
It is a pleasure to thank M. Abbott, G. Arutyunov, R. Borsato, O. Ohlsson Sax, A. Sfondrini and B. Stefanski for interesting discussions and feedback on this work. We would furthermore like to thank the organizers of the workshops \emph{All About $AdS_3$} at ETH Z\"urich and \emph{Holography and Dualities 2016} at Nordita where part of this work was carried out. The work of LW was supported by the ERC Advanced grant No.290456.

\appendix

\section*{Appendix}
\section{The basic one-loop bubble integral}
\label{sec:appendix-integrals}
Here we list the integrated expressions for the basic bubble integral $B^{00}$ that appears in the computations for different configurations of external and internal particles. In the massive sector, where the external particles satisfy $p_+p_-=q_+q_-=1$, we have the following $u$-channel integral
\bea
\label{eq:massive-int}
\textbf{Massive sector:} \qquad
B^{00}_{11}(p,q) = \frac{i}{2\pi} \frac{\log \big(\frac{p_-}{q_-} \big)p_- q_-}{p_-^2-q_-^2}
\eea
For the $t$-channel, where $p=q$, the divergent piece is canceled by pre-factors of external momenta multiplying the integrals. To obtain the $s$-channel one sends $p \rightarrow -p$ and the integral has both real and imaginary terms. The real terms combine into the matrix elements $\ell_i^{(0)}$ determined by the underlying symmetries of the model. 

For the mixed sector we have several distinct integrals depending on the channel. For the $t$-channel both particles have the same mass: Either $p_+p_-=q_+q_-=1$ or $p_+p_-=q_+q_-=0$. The explicit form of the integrals are 
\bea
\label{eq:mixing-int-t}
&&  \textbf{Mixed sector} \textrm{ (t-channel)}\\ \nn
&&
p_+p_-=q_+q_-=1: \\ \nn
&&B^{00}_{11}(p,q) = \frac{i}{2\pi} \frac{\log \big(\frac{p_-}{q_-} \big)p_- q_-}{p_-^2-q_-^2}, \quad B^{00}_{\mu\mu}(p,q) = -\frac{i}{2\pi} \frac{\Big( 2 \log \mu - \log \big(\frac{(p_--q_-)^2}{p_-q_-} \big)\Big)p_- q_-}{(p_--q_-)^3} \\ \nn
&& p_+p_-=q_+q_-=0: \\ \nn
&& B^{00}_{11}(p,q) = \frac{i}{4\pi}, \qquad  B^{00}_{\mu\mu}(p,q) = \frac{i}{2\pi} \frac{1}{\mu^2} \frac{\log \big(\frac{p_-}{q_-} \big)p_- q_-}{p_-^2-q_-^2}
\eea
In the $u$-channel the kinematics are $p_+p_-=1$ and $q_+ q_-=0$ with contributing integrals
\bea
\label{eq:mixing-int-u}
&&  \textbf{Mixed sector} \textrm{ (u-channel):}\qquad B^{00}_{\mu1}  = -\frac{i}{2\pi} \frac{\log\big(\mu\frac{ p_-}{q_-}\big)p_-}{q_-}
\eea
As before we send $p_-\rightarrow - p_-$ to obtain the corresponding $s$-channel integral. 

For the massless sector we again have distinct integrals depending on the channel. For the $t$-channel we have
\bea
\label{eq:massless-int-t}
&&  \textbf{Massless sector} \textrm{ (t-channel)} \\ \nn
&& p_+p_-=q_+q_-=0: \quad 
B^{00}_{\mu\mu}(p,q) =
\frac{i}{2\pi} \frac{1}{\mu^2} \frac{\log \big(\frac{p_\pm}{q_\pm} \big)p_\pm q_\pm}{p_\pm^2-q_\pm^2}
\eea
where $\pm$ indicates the two regimes where the sign of the momentum is positive or negative. Finally, in the $u$-channel we have
\bea
\label{eq:massless-int-u}
&&  \textbf{Massless sector} \textrm{ (u-channel)} \\ \nn
&& p_+p_-=q_+q_-=0: \\ \nn
&& B^{00}_{11}(p,q) = \frac{i}{\pi}\frac{\tanh^{-1} \sqrt{1-\frac{4}{4+p_+q_-}}}{\sqrt{p_+q_-\big(4+p_+q_-\big)}}, \quad B^{00}_{\mu\mu}(p,q) = -\frac{i}{2\pi} \frac{\big(2 \log \mu - \log p_+q_-\big)}{p_+q_-}
\eea
where the momenta satisfy $p>0>q$. Again the $s$-channel is obtained by changing the sign for the in-coming particle. 

\bibliographystyle{JHEP} 

\bibliography{ads2ads3etc}

\end{document}

%% file: dgms.tex
%%
%% This file contains descriptions of feynman diagrams for use with the package feynMP
%% After running the main file, run "mpost diagrams" once to generate these
%%
%%
%% These are for propagators paper with Per.
%% The version in this folder edited 18 May 2011 to add unlabled ones for introduction.
%%

\newsavebox{\feynmanrules}
\sbox{\feynmanrules}{
\begin{fmffile}{diagrams} % I can't seem to make this work using any path but the same one as the document

%%%%%%%%%%%%%%%%
%%  SETTINGS

\fmfset{thin}{0.6pt}  % was 0.7 until v24
%\fmfset{wiggly_len}{5mm}
\fmfset{dash_len}{4pt}
\fmfset{dot_size}{1thick}
\fmfset{arrow_len}{6pt} % you can't use em here, mpost doesn't know what it will be.
%\fmfset{curly_len}{2.5mm}
%\setlength{\unitlength}{1em} % default is =1pt, maybe that's sensible. 72pt = 1in

%%%%%%%%%%%%%%%%%%%%%%%%%%% AdS3 - Scatterings %%%%%%%%%%%%%%%%%%%%%%%%%%%%%%%

%%%% SUNSET

\begin{fmfgraph}(80,40)
\fmfkeep{sunset}
\fmfleft{i}
\fmfright{o}
\fmf{plain,tension=5}{i,v1}
\fmf{plain,tension=5}{v2,o}
\fmf{plain,left,tension=0.4}{v1,v2,v1}
\fmf{plain}{v1,v2}
\fmfdot{v1,v2}
\end{fmfgraph}

%%% DOUBLE BUBBLE
\begin{fmfgraph*}(120,75)
\fmfkeep{doubblebubble}
    \fmfleft{i1,i2,i3}
    \fmfright{o1,o2,o3}
    \fmf{plain}{i1,v1,o1}
    \fmffreeze
    \fmf{phantom}{i2,v2,o2}             
    \fmf{phantom}{i3,v3,o3}
    \fmf{plain,left}{v1,v2,v1}
    \fmf{plain,left}{v2,v3,v2}
    \fmfdot{v1,v2}
\end{fmfgraph*}

\begin{fmfgraph*}(72,25)
\fmfkeep{single}
\fmfleft{in,p1}
\fmfright{out,p2}
\fmfdot{c}
\fmf{dashes_arrow,label=\small{new}}{in,c}
\fmf{dashes_arrow}{c,out}
\fmf{plain_arrow,right, tension=0.8, label=\small{lables}}{c,c}
\fmf{phantom, tension=0.2}{p1,p2}
\end{fmfgraph*}

%%% BUBBLE

\begin{fmfgraph*}(80,40)
\fmfkeep{schannel}
\fmfleft{i1,i2}
\fmfright{o1,o2}
\fmf{plain}{i1,v1}
\fmf{plain}{i2,v1}
\fmf{plain,left=0.5,tension=0.4}{v1,v2}
\fmf{plain,right=0.5,tension=0.4}{v1,v2}
\fmf{plain}{v2,o1}
\fmf{plain}{v2,o2}
\fmfdot{v1,v2}
\end{fmfgraph*}

\begin{fmfgraph*}(80,40)
\fmfkeep{tchannel}
\fmfleft{i1,i2}
\fmfright{o1,o2}
\fmf{plain}{i1,v1,o1}
\fmf{plain}{i2,v2,o2}
\fmf{plain,left=0.5,tension=0.4}{v1,v2}
\fmf{plain,right=0.5,tension=0.4}{v1,v2}
%\fmf{plain}{v2,o1}/
%\fmf{plain}{v2,o2}
\fmfdot{v1,v2}
\end{fmfgraph*}

\begin{fmfgraph*}(80,40)
\fmfkeep{uchannel}
\fmfleft{i1,i2}
\fmfright{o1,o2}
\fmf{plain}{i1,v1}
\fmf{phantom}{v1,o1} % Invisible rubber band
\fmf{plain}{i2,v2}
\fmf{phantom}{v2,o2} % also invisible rubber band
\fmf{plain,left=0.5,tension=0.4}{v1,v2}
\fmf{plain,right=0.5,tension=0.4}{v1,v2}
% These are visible, but have no tension.
\fmf{plain,tension=0}{v1,o2}
\fmf{plain,tension=0}{v2,o1}
\fmfdot{v1,v2}
\end{fmfgraph*}

\begin{fmfgraph*}(80,40)
\fmfkeep{tadpolesix}
\fmfbottom{i1,o1}
\fmftop{i2,o2}
\fmf{plain}{i1,v1,o1}
\fmf{plain}{i2,v1,o2}
\fmf{plain,right=90,tension=0.8}{v1,v1}
%\fmf{plain,label=$(s)$}{v1}
%\fmf{fermion,tension=0}{v1,v2}
\fmfdot{v1}
\end{fmfgraph*}

%%%%%%%%%%%%%%% Two pouint functions

\begin{fmfgraph*}(100,36)
\fmfkeep{bubble}
\fmfleft{in}
\fmfright{out}
\fmfdot{v1}
\fmfdot{v2}
\fmf{plain}{in,v1}
\fmf{plain}{v2,out}
\fmf{plain,left,tension=0.6}{v1,v2}
\fmf{plain,right,tension=0.6}{v1,v2}
\end{fmfgraph*}

%%% TADPOLE

%\begin{fmfgraph*}(72,25)
\begin{fmfgraph*}(100,36)
\fmfkeep{tadpole}
\fmfset{dash_len}{6pt} % this seems to be a local change
\fmfleft{in,p1}
\fmfright{out,p2}
\fmfdot{c}
\fmf{plain}{in,c}
\fmf{plain}{c,out}
\fmf{plain,right, tension=0.8}{c,c}
\fmf{phantom, tension=0.2}{p1,p2}
\end{fmfgraph*}

\begin{fmfgraph*}(100,36)
\fmfkeep{doubletadpole}
\fmfpen{thin}
\fmfleft{i}
\fmfright{o}
\fmf{plain}{i,v,v,o}
\fmf{plain,left=90}{v,v}
\fmfdot{v}
\end{fmfgraph*}

\end{fmffile}

}

%% file: IIB-AdS3-S-matrix-arXiv-v2.bbl
\providecommand{\href}[2]{#2}\begingroup\raggedright\begin{thebibliography}{10}

\bibitem{Wulff:2015mwa}
L.~Wulff, {\it {On integrability of strings on symmetric spaces}},  {\em JHEP}
  {\bf 09} (2015) 115, [\href{http://xxx.lanl.gov/abs/1505.0352}{{\tt
  arXiv:1505.0352}}].

\bibitem{Babichenko:2009dk}
A.~Babichenko, B.~{Stefa\'{n}ski jr.}, and K.~Zarembo, {\it Integrability and
  the {$AdS_3$}/{${CFT}_2$} correspondence},  {\em JHEP} {\bf 03} (2010) 058,
  [\href{http://xxx.lanl.gov/abs/0912.1723}{{\tt arXiv:0912.1723}}].

\bibitem{Sundin:2012gc}
P.~Sundin and L.~Wulff, {\it Classical integrability and quantum aspects of the
  {$AdS_3$} \ensuremath{\times} {$S^3$} \ensuremath{\times} {$S^3$}
  \ensuremath{\times} {$S^1$} superstring},  {\em JHEP} {\bf 10} (2012) 109,
  [\href{http://xxx.lanl.gov/abs/1207.5531}{{\tt arXiv:1207.5531}}].

\bibitem{Cagnazzo:2012se}
A.~Cagnazzo and K.~Zarembo, {\it B-field in {$AdS_3$}/{${CFT}_2$}
  correspondence and integrability},  {\em JHEP} {\bf 11} (2012) 133,
  [\href{http://xxx.lanl.gov/abs/1209.4049}{{\tt arXiv:1209.4049}}].

\bibitem{Sundin:2013uca}
P.~Sundin and L.~Wulff, {\it {The low energy limit of the AdS(3) x S(3) x M(4)
  spinning string}},  {\em JHEP} {\bf 10} (2013) 111,
  [\href{http://xxx.lanl.gov/abs/1306.6918}{{\tt arXiv:1306.6918}}].

\bibitem{Wulff:2014kja}
L.~Wulff, {\it {Superisometries and integrability of superstrings}},  {\em
  JHEP} {\bf 05} (2014) 115, [\href{http://xxx.lanl.gov/abs/1402.3122}{{\tt
  arXiv:1402.3122}}].

\bibitem{Beisert:2010jr}
N.~Beisert et~al., {\it Review of {AdS}/{CFT} integrability: An overview},
  {\em Lett. Math. Phys.} {\bf 99} (2012) 3--32,
  [\href{http://xxx.lanl.gov/abs/1012.3982}{{\tt arXiv:1012.3982}}].

\bibitem{Borsato:2014hja}
R.~Borsato, O.~Ohlsson~Sax, A.~Sfondrini, and B.~Stefanski, {\it {The complete
  $AdS_3\times S^3 \times T^4$ worldsheet S matrix}},  {\em JHEP} {\bf 1410}
  (2014) 66, [\href{http://xxx.lanl.gov/abs/1406.0453}{{\tt arXiv:1406.0453}}].

\bibitem{Borsato:2013hoa}
R.~Borsato, O.~O. Sax, A.~Sfondrini, J.~Stefanski, Bogdan, and A.~Torrielli,
  {\it {Dressing phases of AdS3/CFT2}},  {\em Phys.Rev.} {\bf D88} (2013)
  066004, [\href{http://xxx.lanl.gov/abs/1306.2512}{{\tt arXiv:1306.2512}}].

\bibitem{Sundin:2013ypa}
P.~Sundin and L.~Wulff, {\it Worldsheet scattering in {$AdS_3$}/{${CFT}_2$}},
  {\em JHEP} {\bf 07} (2013) 007,
  [\href{http://xxx.lanl.gov/abs/1302.5349}{{\tt arXiv:1302.5349}}].

\bibitem{Engelund:2013fja}
O.~T. Engelund, R.~W. McKeown, and R.~Roiban, {\it {Generalized unitarity and
  the worldsheet $S$ matrix in $AdS_n \times S^n \times M^{10-2n}$}},  {\em
  JHEP} {\bf 1308} (2013) 023, [\href{http://xxx.lanl.gov/abs/1304.4281}{{\tt
  arXiv:1304.4281}}].

\bibitem{Abbott:2013mpa}
M.~C. Abbott, {\it {The $AdS_{3}\times S^{3}\times S^{3}\times S^{1}$
  Hern\'andez -– L\'opez phases: a semiclassical derivation}},  {\em J.
  Phys.} {\bf A46} (2013) 445401,
  [\href{http://xxx.lanl.gov/abs/1306.5106}{{\tt arXiv:1306.5106}}].

\bibitem{Bianchi:2014rfa}
L.~Bianchi and B.~Hoare, {\it {$AdS_3 \times S^3 \times M^4$ string S-matrices
  from unitarity cuts}},  {\em JHEP} {\bf 08} (2014) 097,
  [\href{http://xxx.lanl.gov/abs/1405.7947}{{\tt arXiv:1405.7947}}].

\bibitem{Roiban:2014cia}
R.~Roiban, P.~Sundin, A.~Tseytlin, and L.~Wulff, {\it {The one-loop worldsheet
  S-matrix for the $AdS_n\times S^n\times T^{10-2n}$ superstring}},  {\em JHEP}
  {\bf 1408} (2014) 160, [\href{http://xxx.lanl.gov/abs/1407.7883}{{\tt
  arXiv:1407.7883}}].

\bibitem{Sundin:2015uva}
P.~Sundin and L.~Wulff, {\it {The AdS$_{n} \times$ S$^{n} \times$ T$^{10-2n}$
  BMN string at two loops}},  {\em JHEP} {\bf 11} (2015) 154,
  [\href{http://xxx.lanl.gov/abs/1508.0431}{{\tt arXiv:1508.0431}}].

\bibitem{Borsato:2016kbm}
R.~Borsato, O.~O. Sax, A.~Sfondrini, and B.~Stefanski, {\it {On the spectrum of
  $AdS_3 \times S^3 \times T^4$ strings with Ramond-Ramond flux}},
  \href{http://xxx.lanl.gov/abs/1605.0051}{{\tt arXiv:1605.0051}}.

\bibitem{Abbott:2015pps}
M.~C. Abbott and I.~Aniceto, {\it {Massless L\"uscher Terms and the Limitations
  of the AdS3 Asymptotic Bethe Ansatz}},
  \href{http://xxx.lanl.gov/abs/1512.0876}{{\tt arXiv:1512.0876}}.

\bibitem{Sundin:2014ema}
P.~Sundin and L.~Wulff, {\it {One- and two-loop checks for the $AdS_3\times
  S^3\times T^4$ superstring with mixed flux}},  {\em J. Phys.} {\bf A48}
  (2015), no.~10 105402, [\href{http://xxx.lanl.gov/abs/1411.4662}{{\tt
  arXiv:1411.4662}}].

\bibitem{Wulff:2013kga}
L.~Wulff, {\it {The type II superstring to order $\theta^4$}},  {\em JHEP} {\bf
  1307} (2013) 123, [\href{http://xxx.lanl.gov/abs/1304.6422}{{\tt
  arXiv:1304.6422}}].

\bibitem{Berenstein:2002jq}
D.~E. Berenstein, J.~M. Maldacena, and H.~S. Nastase, {\it Strings in flat
  space and {PP} waves from $\mathcal{N}\!=4$ super {Yang Mills}},  {\em JHEP}
  {\bf 04} (2002) 013, [\href{http://xxx.lanl.gov/abs/hep-th/0202021}{{\tt
  hep-th/0202021}}].

\bibitem{Bianchi:2015vgw}
L.~Bianchi and M.~S. Bianchi, {\it {On the scattering of gluons in the GKP
  string}},  {\em JHEP} {\bf 02} (2016) 146,
  [\href{http://xxx.lanl.gov/abs/1511.0109}{{\tt arXiv:1511.0109}}].

\bibitem{Zarembo:2009au}
K.~Zarembo, {\it Worldsheet spectrum in {$AdS_4$}/{${CFT}_3$} correspondence},
  {\em JHEP} {\bf 04} (2009) 135,
  [\href{http://xxx.lanl.gov/abs/0903.1747}{{\tt arXiv:0903.1747}}].

\bibitem{OhlssonSax:2011ms}
O.~Ohlsson~Sax and B.~{Stefa\'{n}ski jr.}, {\it Integrability, spin-chains and
  the {$AdS_3$}/{${CFT}_2$} correspondence},  {\em JHEP} {\bf 08} (2011) 029,
  [\href{http://xxx.lanl.gov/abs/1106.2558}{{\tt arXiv:1106.2558}}].

\bibitem{Sax:2012jv}
O.~Ohlsson~Sax, j.~Stefanski, Bogdan, and A.~Torrielli, {\it {On the massless
  modes of the AdS3/CFT2 integrable systems}},  {\em JHEP} {\bf 1303} (2013)
  109, [\href{http://xxx.lanl.gov/abs/1211.1952}{{\tt arXiv:1211.1952}}].

\bibitem{Hoare:2013lja}
B.~Hoare, A.~Stepanchuk, and A.~Tseytlin, {\it {Giant magnon solution and
  dispersion relation in string theory in $AdS_3\times S^3\times T^4$ with
  mixed flux}},  {\em Nucl.Phys.} {\bf B879} (2014) 318--347,
  [\href{http://xxx.lanl.gov/abs/1311.1794}{{\tt arXiv:1311.1794}}].

\bibitem{Lloyd:2014bsa}
T.~Lloyd, O.~Ohlsson~Sax, A.~Sfondrini, and J.~Stefanski, Bogdan, {\it {The
  complete worldsheet S matrix of superstrings on $AdS_3\times S^3 \times T^4$
  with mixed three-form flux}},  {\em Nucl.Phys.} {\bf B891} (2015) 570--612,
  [\href{http://xxx.lanl.gov/abs/1410.0866}{{\tt arXiv:1410.0866}}].

\bibitem{Borsato:2013qpa}
R.~Borsato, O.~Ohlsson~Sax, A.~Sfondrini, B.~Stefański, and A.~Torrielli, {\it
  {The all-loop integrable spin-chain for strings on AdS$_3 \times S^3 \times
  T^4$: the massive sector}},  {\em JHEP} {\bf 1308} (2013) 043,
  [\href{http://xxx.lanl.gov/abs/1303.5995}{{\tt arXiv:1303.5995}}].

\bibitem{Klose:2007rz}
T.~Klose, T.~McLoughlin, J.~A. Minahan, and K.~Zarembo, {\it World-sheet
  scattering in {$AdS_5\times S^5$} at two loops},  {\em JHEP} {\bf 08} (2007)
  051, [\href{http://xxx.lanl.gov/abs/0704.3891}{{\tt arXiv:0704.3891}}].

\bibitem{Borsato:2014exa}
R.~Borsato, O.~Ohlsson~Sax, A.~Sfondrini, and B.~Stefanski, {\it {Towards the
  All-Loop Worldsheet S Matrix for $AdS_3\times S^3\times T^4$}},  {\em Phys.
  Rev. Lett.} {\bf 113} (2014), no.~13 131601,
  [\href{http://xxx.lanl.gov/abs/1403.4543}{{\tt arXiv:1403.4543}}].

\bibitem{Abbott:2012dd}
M.~C. Abbott, {\it Comment on strings in {$AdS_3$} \ensuremath{\times} {$S^3$}
  \ensuremath{\times} {$S^3$} \ensuremath{\times} {$S^1$} at one loop},  {\em
  JHEP} {\bf 02} (2013) 102, [\href{http://xxx.lanl.gov/abs/1211.5587}{{\tt
  arXiv:1211.5587}}].

\bibitem{Beccaria:2012kb}
M.~Beccaria, F.~Levkovich-Maslyuk, G.~Macorini, and A.~Tseytlin, {\it Quantum
  corrections to spinning superstrings in {$AdS_3$} \ensuremath{\times} {$S^3$}
  \ensuremath{\times} {$M^4$}: determining the dressing phase},  {\em JHEP}
  {\bf 04} (2013) 006, [\href{http://xxx.lanl.gov/abs/1211.6090}{{\tt
  arXiv:1211.6090}}].

\bibitem{Borsato:2012ss}
R.~Borsato, O.~Ohlsson~Sax, and A.~Sfondrini, {\it {All-loop Bethe ansatz
  equations for AdS3/CFT2}},  {\em JHEP} {\bf 1304} (2013) 116,
  [\href{http://xxx.lanl.gov/abs/1212.0505}{{\tt arXiv:1212.0505}}].

\bibitem{Borsato:2015mma}
R.~Borsato, O.~Ohlsson~Sax, A.~Sfondrini, and B.~Stefanski, {\it {The
  $\mathrm{AdS}_3\times \mathrm{S}^3\times \mathrm{S}^3\times\mathrm{S}^1$
  worldsheet S matrix}},  {\em J. Phys.} {\bf A48} (2015), no.~41 415401,
  [\href{http://xxx.lanl.gov/abs/1506.0021}{{\tt arXiv:1506.0021}}].

\bibitem{Rughoonauth:2012qd}
N.~Rughoonauth, P.~Sundin, and L.~Wulff, {\it Near-{BMN} dynamics of the
  {$AdS_3$} \ensuremath{\times} {$S^3$} \ensuremath{\times} {$S^3$}
  \ensuremath{\times} {$S^1$} superstring},  {\em JHEP} {\bf 07} (2012) 159,
  [\href{http://xxx.lanl.gov/abs/1204.4742}{{\tt arXiv:1204.4742}}].

\end{thebibliography}\endgroup
